\documentclass[journal=jacsat,manuscript=communication]{achemso}
\usepackage{chemformula} 
\usepackage[T1]{fontenc} 
\usepackage{titlecaps}
\usepackage{longtable}
\usepackage{wrapfig}
\usepackage{color}

\author{Md Omar Faruque}
\author{Suchona Akter}
\author{Dil K. Limbu}
\author{Kathleen Kilway}
\author{Zhonghua Peng$^*$}
\author{Mohammad R. Momeni}
\email{pengz@umkc.edu, mmomenitaheri@umkc.edu}
\affiliation{Division of Energy, Matter and Systems, School of Science and Engineering, University of Missouri $-$ Kansas City, Kansas City 64110, MO United States}

\title[An \textsf{achemso} demo]
{High throughput screening, crystal structure prediction, and carrier mobility calculations of organic molecular semiconductors as hole transport layer materials in perovskite solar cells }

\abbreviations{}

\begin{document}

\begin{abstract}
Using a representative translational dimer model, high throughput calculations are implemented for fast screening of a total of 74 diacenaphtho-extended heterocycle (DAH) derivatives as hole transport layer (HTL) materials in perovskite solar cells (PVSCs). Different electronic properties, including band structures, band gaps, and band edges compared to methylammonium and formamidinium lead iodide perovskites, along with reorganization energies, electronic couplings, and hole mobilities are calculated in order to decipher the effects of different parameters, including the polarity, steric and $\pi$-conjugation, as well as the presence of explicit hydrogen bond interactions on the computed carrier mobilities of the studied materials. Full crystal structure predictions and hole mobility calculations of the top candidates resulted in some mobilities exceeding 10 cm$^2$/V.s, further validating the employed translational dimer model as a robust approach for inverse design and fast high throughput screening of new HTL organic semiconductors with superior properties. The studied models and simulations performed in this work are instructive in designing next-generation HTL materials for higher-performance PVSCs.
\end{abstract}

\section{1. Introduction}  \label{sec1}
\textbf{1. Introduction}\newline
Perovskite solar cells (PVSCs) are among the most exciting and promising new solar cell technologies hailed as the future of solar power \cite{kojima2009organometal, lee2012efficient, jeon2015compositional, yang2015high, chen2015efficient, li2016vacuum, arora2017perovskite, yang2017iodide, jeon2018fluorene}. The first PVSC was reported in 2009 with a power conversion efficiency (PCE) lower than 4\% \cite{kojima2009organometal}. Since then, the efficiency of PVSCs has skyrocketed to over 25\% \cite{min2021perovskite}, surpassing multi-crystalline Si-based solar cells, making them the most appealing new technology for solar-electricity conversion \cite{green2019solar}. A typical PVSC is comprised of a central light-absorbing perovskite layer sandwiched with two electrodes, a cathode and an anode, which collect photogenerated electrons and holes, respectively, from the perovskite. To facilitate unidirectional charge transfer and collection, an electron-transporting layer (ETL) is introduced between the perovskite and the cathode, while a hole-transporting layer (HTL) is inserted between the perovskite and the anode. 

The HTL plays a critical role in device performance in terms of both its efficiency and long-term stability \cite{gratia2015methoxydiphenylamine, huang2016dopant, lee2017green, kim2017engineering, park20152, bi2016high}. Some of the most important properties of an ideal HTL material include (1) high intrinsic hole mobility (> 10$^{-2}$ cm$^2$/V$\cdot$s) without the need for any doping; (2) good valence and conduction band alignments with perovskites; (3) high chemical, thermal, and photostability; and (4) low cost with convenient synthesis from readily available source materials.\cite{gratia2015methoxydiphenylamine, huang2016dopant, lee2017green, kim2017engineering, park20152, bi2016high, calio2016hole} Despite advances in the development of a variety of HTL materials, 2,2$^{\prime}$,7,7$^{\prime}$-tetrakis(N, N-di-p-methoxyphenylamine)-9,9$^{\prime}$-spirobifluorene (spiro-MeOTAD) remains the HTL of choice with the best-reported performances \cite{min2021perovskite, kim2020high, bakr2017advances, pham2020development}. Yet, spiro-MeOTAD is expensive and is a poor conductor with low hole mobility ($\approx$10$^{-5}$ cm$^2$/V$\cdot$s) and thus requires a complicated doping process that is difficult to control \cite{bakr2017advances}. The dopants are often ions that attract moisture (Li-TFSI, Na-TFSI, Ag-TFSI, etc.), degrading the perovskite layer \cite{bakr2017advances}. Realizing that PVSCs with spiro-MeOTAD as the HTL will not be commercially viable, there have been tremendous research efforts devoted to developing new HTL materials \cite{calio2016hole, bakr2017advances, pham2020development, urieta2018hole, singh2019review}. While inorganic HTLs have also been explored \cite{singh2019review}, they often react with the perovskite and the electrode, compromising device stability \cite{kim2020high}. 

On the other end of alternative HTL materials are organic molecular crystals (OMCs) with rigid, planar, or near planar, $\pi$-conjugated cores.\cite{li2018highly, li2014unusually, li2016spin} 
OMCs with high intrinsic carrier mobilities have made themselves indispensable not only to PVSCs, the focus of the current study, but also to other applications in electronics and photonics, including organic field-effect transistors\cite{sci_303_1644, prl_93_086602} as well as organic light-emitting diodes\cite{apl_51_913} and organic photovoltaics,\cite{CHAMBERLAIN198347} among others. In all the above-mentioned applications, high carrier mobility OMCs are essential for realizing high-performance devices. Overall, high charge carrier mobilities in OMCs can arise when long-range order is achieved through intermolecular $\pi$-stacking of the rigid core, forming columnar structures. However, typical OMCs such as tetracene and pentacene can be easily oxidized in air to form endoperoxides \cite{kalb2008oxygen} or undergo dimerization, especially under radiation.\cite{nguyen2015density, dong2019photochemistry} Here, our theoretical calculations are focused on diacenaphtho-extended [1,2-b:1$^{\prime}$,2$^{\prime}$-d] heterocycles (DAHs) which can potentially satisfy all above-mentioned properties for HTLs. Specifically, DAHs have a planar rigid core with $\pi$-extension much longer in one direction, similar to acenes. All aromatic rings are angularly fused, giving the system high chemical, thermal, and photostability. Moreover, the central heterocycle introduces polarity into the $\pi$ system, which may increase the cofacial $\pi-\pi$ stacking interactions. All these features point to possible strong $\pi-\pi$ stacking among DAH cores with limited rotational freedom, leading to a high charge carrier mobility. When the central heterocycle is a thiophene ring, the corresponding DAT core can be conveniently synthesized in one straightforward step from the readily available inexpensive acenaphthene \cite{clapp1939reaction, li2014unusually, li2016spin, li2018highly}, which distinguishes it from the complicated multistep synthesis of nearly all other OMCs with a similar number of sp$^2$ carbon atoms. 
\begin{figure*}[!t]
\centering
\includegraphics[width=0.8\linewidth]{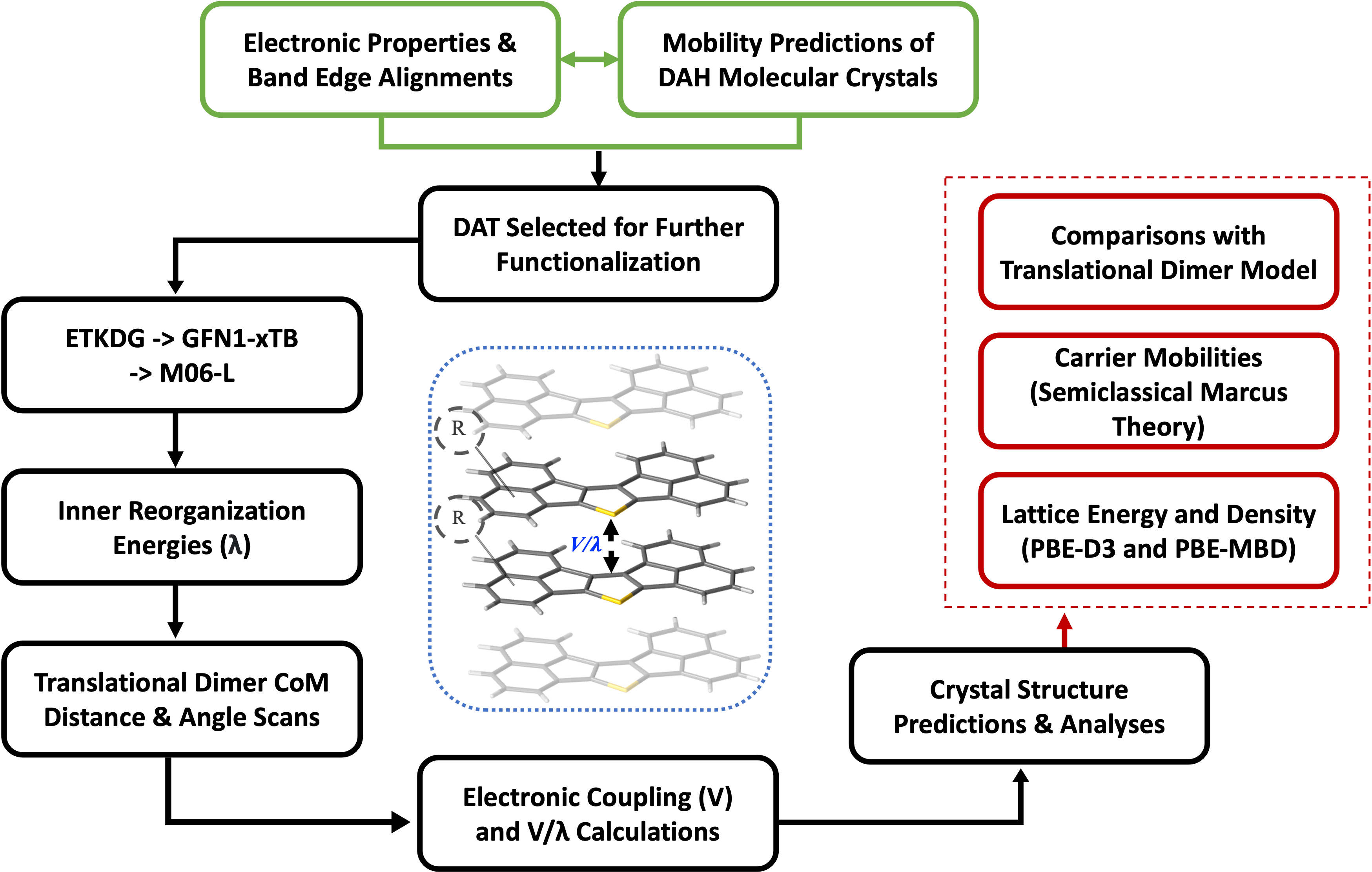}
\caption{High-throughput screening workflow adapted in this work using the translational dimer model. Different functional groups are attached in the positions shown in the Supporting Information (SI) Figure S4. The dimers are screened using the ratios of the electronic coupling over reorganization energy (i.e., V/$\lambda$). For the dimers with the highest $V/\lambda$ values, crystal structures are predicted with their mobilities calculated using the semiclassical Marcus theory.}
\label{workflow}
\end{figure*}
Similarly, having 1-phenyl-1H-pyrrole and 1,1-diphenyl-1-H-silole as the central heterocycle, one can obtain the corresponding DAP and DAS core structures, respectively. 

In this work, we aim to not only perform high-throughput screening calculations for potential DAH-based OMCs but also to provide fundamental new insights into structure-property relationships in these materials, for example, by probing the effects of molecular packing on the calculated electronic couplings and carrier mobilities, among others. First, the electronic properties and hole mobilities of DAT, DAP, and DAS core materials as potential HTL materials were calculated using their experimentally reported crystal structures.\cite{nagasaka2011diacenaphtho, adams2004diacenaphtho} Specifically, density functional theory (DFT) calculated electronic band structures, band gaps, and band edges, along with reorganization energies, electronic couplings, and hole mobilities, were used to establish trends and provide fundamental understandings of charge carrier mobilities in these materials. After initial screening of the considered DAH cores, DAT was selected for systematic functionalization, with its mobilities screened by utilizing a translational dimer model as an efficient proxy. This initial screening resulted in several structures with superior hole mobilities compared to parent DAT. To design novel organic semiconductors with desired properties, the solid-state packing of the potential candidates must be predicted. Extensive theoretical research is being conducted to search the infinitely vast chemical space of potential OMCs with high carrier mobilities using a combination of crystal structure predictions (CSPs) and hole mobility calculations.\cite{schmidt2021cgd, salerno2019influence, jpcc_113_6821, pccp_18_21371, fd_174_281, deng2004predictions, sokolov2011computational, jpcb_109_1849, cgd_6_1697} For CSP, we implemented an effective workflow leveraging machine learning (see Figure \ref{workflow} for the details of our employed approach). The mobilities of the predicted crystal structures were then calculated and compared to those obtained via the adapted translational dimer approach. This work is structured as follows: our theory details are presented in the next section, section 3 contains the results and their corresponding discussions, followed by conclusions and future outlooks.

\section{2. Models and Simulation Details} \label{sec2}
\textbf{2. Models and Simulation Details}\newline
\subsection{2.1 Periodic Electronic Structure Calculations} \label{sec2.1}
\begin{figure*}[!h]
	\centering
	\includegraphics[width=0.9\linewidth]{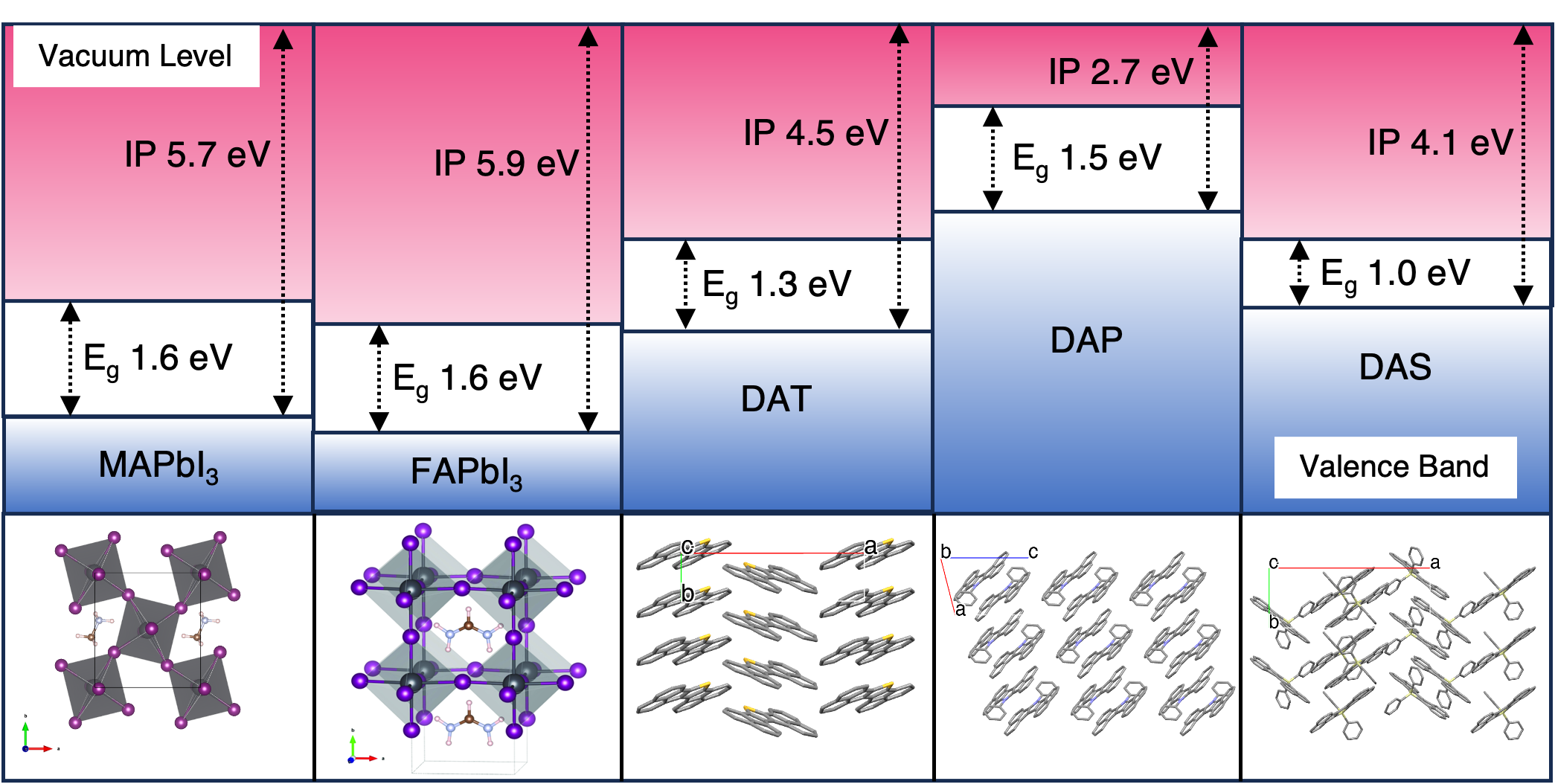}
	\caption{PBE-D3 calculated vacuum-aligned band edges of DAT, DAP, and DAS compared to MAPbI$_3$ and FAPbI$_3$ perovskites. Computed band gaps (E$_g$ in eV) and ionization potentials (IPs in eV) are also given.}
	\label{electronic}
\end{figure*}

\textbf{2.1 Periodic Electronic Structure Calculations}\newline
The experimental crystal structures of all studied DAH cores were obtained from the Cambridge Crystallographic Data Center (CCDC).\cite{nagasaka2011diacenaphtho, adams2004diacenaphtho} The DAT and DAP crystal structures were used as is while in the case of DAS, solvent molecules were removed (Figure \ref{electronic}). All structure minimizations and electronic property calculations were performed using the Vienna \textit{ab initio} simulation package--VASP.\cite{Kresse1993,Kresse1994,Kresse1996,Kresse19962} 

First, a careful benchmark of minimized atomic positions and cell vectors was performed, without imposing any constraints, using different exchange-correlation functionals starting from the widely employed Perdew--Burke--Ernzenhof (PBE)\cite{Perdew:1996} functional as well as the revised PBE (RPBE),\cite{Hammer:1999} and PBE for solids (PBEsol).\cite{Perdew:2008} 
Organic molecules in all studied solid-state crystals are $\pi$-stacked via van der Waals interactions, and hence, the correct description of these interactions is of utmost importance. Therefore, a series of dispersion correction schemes in the form of PBE functional with Becke-Johnson and zero-damped dispersion correction\cite{Grimme:2010}, Tkatchenko-Scheffler (PBE-TS) \cite{tkatchenko2009accurate}, and many-body dispersion (PBE-MBD) \cite{ambrosetti2014long,tkatchenko2012accurate} schemes were employed. The results of these benchmarks are provided in the Supporting Information (SI) Table S1. Expectedly, all pure GGA functionals with no dispersion corrections were found to grossly overestimate the cell vectors and the center of mass (CoM) distances due to not accounting for these interactions. Adding Grimme's zero-damped D3 dispersion correction\cite{Grimme:2010} to the PBE functional resulted in structural data that closely matched those of experiments. 
Therefore, PBE-D3 with zero-damping (PBE-D3 hereafter) was used throughout this work unless otherwise stated.

Interactions between electrons and ions were described by Projector Augmented Wave (PAW) potentials \cite{PAW1, PAW2} with the energy cutoff of 500 eV. Gaussian smearing was adopted in all atomic position and cell vector minimizations, as well as band structure calculations with a smearing width of 0.05 eV, while the tetrahedron method was used for the calculated projected density of states (pDOS).\cite{sholl2011density} The convergence criteria were set to 10$^{-5}$ for SCF calculations and $10^{-6}$ for electronic property calculations. The $k$--point mesh in the Monkhorst--Pack scheme was set to 2$\times$6$\times$1, 4$\times$3$\times$3, and 1$\times$3$\times$1 in the SCF part for DAT, DAP, and DAS, respectively and twice denser in the following electronic property calculations; see SI Figure S1 for the $k$--point paths. Spin-polarized calculations were performed for all systems. The PBE-D3 calculated band structures and pDOSs of all studied systems are given in SI Figure S3. 

Vacuum-aligned band edges were calculated according to the method proposed in Ref. \citenum{walsh2015principles} for all studied core OMCs along with two of the most widely used perovskite layer materials: methylammonium lead iodide (MAPbI$_3$) and formamidinium lead iodide (FAPbI$_3$); see Figure \ref{electronic}. 
Briefly, slabs were generated for all systems with at least 15~\AA~of vacuum in the direction that is normal to the slab surface. After a ground state DFT-D3 calculation with VASP, the vacuum potentials were calculated for each of the considered systems. The vacuum potentials were aligned to a common vacuum level at 0 V, and the ionization potentials (IP) were calculated as:

\begin{equation}
IP=  E_{VAC} - E_{F}
\end{equation}
where $E_{VAC}$ and $E_{F}$ are the vacuum potential and Fermi energy respectively.

\subsection{2.2 Hole Mobility Calculations}\label{sec2.2}
\textbf{2.2 Hole Mobility Calculations}\newline
Information about charge mobility and its relation with different packing motifs can provide important design guidelines. The standard semiclassical Marcus theory represents an excellent tool for studying carrier mobility in solid-state molecular crystals with its practical applications shown by Deng and Goddard for similar organic semiconductor crystals based on oligoacenes \cite{deng2004predictions}. According to this theory, one can estimate carrier mobility rates (W$_i$) between near neighbor units as:
\begin{equation}
W_i=  \frac{|V_i|^2}{\hbar} \sqrt{\frac{\pi}{\lambda k_BT}} ~exp {(-\frac{\lambda}{4 k_B T})}
\end{equation}
where $k_B$ and T are the Boltzmann constant and temperature (300 K), |V$_i$| represents electronic couplings between a representative central unit with its near neighbors in the crystal (see SI Figure S7), and the decaying exponential corresponds to the energy barrier for the charge transport between two consecutive units which is dominated by $\lambda_{+/-}$ (the cationic/anionic reorganization energy corresponding to the transport of hole/electron, respectively).
Similar to our previous study on $\pi$-stacked layered MOFs,\cite{dell2022modeling} the $\lambda$ term was calculated using the standard four-point scheme and molecular monomers at the M06-2X/cc-pVTZ//M06L/cc-pVTZ level of theory in Gaussian 16.\cite{g16} To ensure accuracy, we performed benchmarks on similar oligoacenes with available reference data. For example, our M06-2X/cc-pVTZ//M06L/cc-pVTZ calculated $\lambda_+$ values of 0.120 eV for tetracene and 0.101 eV for pentacene agree very well with their corresponding experimentally deduced values of 0.118 eV and 0.099 eV \cite{ostroverkhova2013handbook}. Electronic coupling, which expresses the ease of charge transfer between two neighboring units, is well known to be estimated to a very good approximation for hole transport as half of the splitting of HOMO for different dimers. \cite{cornil2001interchain, kashimoto2018evolution} The intermolecular electronic couplings for hole transport in all adjacent dimers were calculated using the M06-2X functional and the cc-pVTZ basis set. Using calculated mobility rates, absolute hole mobilities ($\mu$) can be computed according to the Einstein equation:
\begin{equation}
\mu = \frac{(e\times D)}{(k_B T)}
\end{equation}
where $e$ is the electronic charge, and $D$ is the diffusion coefficient written as a sum over all the near neighbors in the crystal:
\begin{equation}
D=  \frac{1}{2n} \sum_i r^2 W_i  P_i
\end{equation}
In the above equation, $n$ is the dimensionality (3 in our case), $r$ is the distance from the CoM of the representative unit to its i-th near neighbor, and P$_i$ corresponds to the probability of hopping between different dimer units:
\begin{equation}
P_i = \frac{W_i}{\sum_i W_i }
\end{equation}
In-house scripts were developed and used for computational high-throughput screening of $\lambda$, V$_i$, W$_i$, D, and finally $\mu$ for each studied crystal according to the semiclassical Marcus theory discussed above.

\subsection{2.3 High-Throughput Screening Using the Translational Dimer Model}\label{sec2.3}
\textbf{2.3 High-Throughput Screening Using the Translational Dimer Model}\newline
Various classes of functional groups were chosen in this work to test the effects of bulkiness, polarity, and $\pi$-extension on calculated electronic properties and hole mobilities. Guided by the employed procedure to synthesize these materials, the functional groups were substituted in symmetric peripheral positions, as shown in the SI Figure S4. The full list of all functional groups is provided in the SI Figure S5. A modified version of the \textit{chem\_tools} code\cite{Turcani_2020} was used to generate the substituted DAT monomers. Other than functionalization, homo and heterocyclic aromatic ring fused $\pi$ extension was also used. These ring-extended structures are shown in SI Figure S6.

The generated systems were first optimized using the built-in ETKDG\cite{riniker2015etkdg} optimizer in the RDkit\cite{RDKit} code. Subsequently, to explore the lowest energy conformers, molecular dynamics simulations were performed at 700 K with a timestep of 1.0 fs and a total of 2000 steps with conformers sampled at each timestep. For these simulations, the isolated molecule, as obtained from the previous step, was placed at the center of a cubic box with side lengths of 25 \AA. The semi-empirical tight-binding GFN1-xTB \cite{grimme2017robust} method as implemented in CP2K\cite{kuhne2020cp2k} was used for these simulations. The lowest energy conformers were further optimized using M06L/cc-pVTZ before calculating the reorganization energies with the M06-2X/cc-pVTZ level of theory as discussed above.

For all studied systems, the near neighbor unit is a translational dimer with the CoM distances of 3.9~\AA, 3.8~\AA, and 5.8~\AA~for DAT, DAP, and DAS, respectively; see the SI Figure S7. Previous studies have indicated that the near neighbor translational dimer units in similar systems exhibit the highest electron transfer integrals.\cite{schmidt2021cgd, salerno2019influence}. Therefore, it can be assumed with a good approximation that the translational dimer constitutes the dominant transport channel, and for accelerated screening, it can be used as a representative of the entire molecular crystal's mobility. Our calculations show that using the nearest translational dimer unit in DAT, the calculated hole mobility (0.80 cm$^2$/V.s) closely matches that of the entire crystal (0.73 cm$^2$/V.s) where all adjacent neighbors are considered. Therefore, the translational dimer model was employed for high-throughput screening of the electronic coupling and hole mobility of all functionalized DAT systems considered in this work. 

To locate the lowest energy dimers of all modified DATs, a comprehensive distance and angle potential energy surface (PES) scan was performed from 3--10~\AA~ with 0.1~\AA~ increments and from 0$^\circ$ to 180$^\circ$ angle with 15$^\circ$ increments, respectively. The GFN1-xTB method was chosen for performing these high-throughput PES scans as an inexpensive and accurate method. GFN-xTB has performed very well in generating accurate structures close to the experiment.\cite{schmidt2021cgd} In our systems, among different variations of the GFN-xTB methods, the GFN1-xTB variant was found to produce the closest structures to the experiment  (SI Table S3). For DAT, GFN1-xTB predicted the lowest energy dimer at 3.4~\AA, compared to the PBE-MBD lowest energy dimer with a CoM distance of 3.6~\AA~(SI Table S3). All screened dimers were then ranked based on their relative energies, with the top 20 isomers selected for calculating averaged electronic couplings, V/$\lambda$, and mobilities. Finally, to verify the accuracy of our rankings using the translational dimer model, the crystal structures of the top-performing candidates were predicted, with their carrier mobilities calculated using the semi-classical Marcus theory.

\subsection{2.4 Crystal Structure Predictions}\label{sec2.4}
\textbf{2.4 Crystal Structure Predictions}\newline
The crystal structures of the top four novel structures from both functionalized and ring-extended DATs with the highest V/$\lambda$ values were predicted. First, random crystal structures were generated using the Genarris 2.0 package using a modified CSP workflow.\cite{tom2020genarris} First, initial structures were generated based on the machine-learned model predicted volume as implemented in Genarris 2.0. During structure generation, the predicted volume was used as the mean volume, but a six-fold prediction error was used as the standard deviation instead of the three-fold as implemented in Genarris. This ensured the inclusion of a diverse landscape of initial structures, which were even larger than the structure pools generated in the original Genarris studies.\cite{tom2020genarris,li2018genarris} Next, all generated crystal structures were clustered based on the radial symmetry function (RSF) descriptor and the affinity propagation (AP) machine learning algorithm. Then, they were subjected to dispersion-included PBE-D3 calculations in VASP. All single-point energy calculations in this stage were performed at the $\Gamma$ point. Later, after selecting the lowest energy structures from each cluster through a second AP clustering, their geometry was optimized by minimizing both atomic positions and cell vectors. PBE-D3 was used for these optimizations, except for the M6 system, for which PBE-MBD was used for more accurate rankings.

In all calculations, the cutoff for plane wave basis sets was set to 500 eV, and a density of at least 0.05 \AA $^{-1}$ was employed for the $k$-point grids. The crystal interaction energies ($E_{int}$) were calculated as $E_{int}$ = $E_{crys}$ – ($n~\times$ $E_{mol}$) where $E_{crys}$ and $E_{mol}$ are calculated total energies of the crystal structure and isolated molecule, respectively, with $n$ corresponding to the number of molecules in the unit cell. Finally, the calculated $E_{int}$ vs. densities were analyzed to identify the top candidates for each unknown system. Translational motifs within the final predicted crystal structures were identified with an in-house code. To ensure the accuracy of our employed methodology, we performed validation tests for the experimentally reported crystal structure of the parent DAT core\cite{musgrave2004diacenaphtheno} as reference. The performance of this validation test is shown in the SI Figure S10, where the experimental structure of DAT was successfully predicted.

\section{3. Results and Discussion} \label{sec3}
\textbf{3. Results and Discussion}\newline
\subsubsection{3.1 Electronic Properties of DAH Cores} \label{sec3.1}
\textbf{3.1 Electronic Properties of DAH Cores}\newline
PBE-D3 calculated band gaps for DAT, DAP, and DAS are 1.3 eV, 1.5 eV, and 1.0 eV, respectively (see SI Figure S3 for all calculated band structures and pDOSs). The DAT core structure shows a direct band gap, whereas DAP and DAS illustrate an indirect band gap. Based on these computed band gaps, which provide a lower bound to more accurate HSE06 data (1.7 eV and 2.2 eV for DAT and DAP, respectively), all studied materials can be classified as semiconductors. Using our calculated band structures, the effective masses for holes ($m_h^*$), which are proportional to the valence band curvature and the valence band maximum (VBM) bandwidths, are calculated (see Table \ref{table1}). 
\begin{table}[!h]
\centering
\resizebox{0.99\linewidth}{!}{
    	\begin{tabular}{ccccccc} \hline        
    	Systems & CoM (\AA) & VBM BW (eV) & $m_h^*$ & $\lambda$ (eV) & $\mu$ (${cm^2/Vs}$) \\ \hline 
    	DAT & 3.91 & 0.66 & 4.82 & 0.30 & 0.73 \\ 
    	DAP & 3.80 & 0.11 & 7.44 & 0.31 & 0.11 \\  
    	DAS & 5.81 & 0.07 & 15.97 & 0.40 & 0.16 \\ \hline
	\end{tabular}}
	\caption{Calculated near neighbor center of mass distance (CoM), VBM bandwidth (VBM BW), hole effective mass $(m_h^*$ in the unit of electron rest mass $m_0$), reorganization energy ($\lambda$) and hole mobility ($\mu$) of DAT, DAP, and DAS.}
	\label{table1}
\end{table}
Qualitatively, a less curved band results in a heavier $m_h^*$ leading to a slower charge transport. For DAT, the calculated $m_h^*$ is the lowest (4.82$m_0$) compared to DAP and DAS (7.44$m_0$ and 15.97$m_0$). Also, a large VBM bandwidth of 0.66 eV is obtained for DAT compared to DAP and DAS, with significantly smaller bandwidths of 0.11 eV and 0.07 eV, respectively. Based on band structure analyses, one can predict the hole mobility of DAT to be higher than the other two systems, making it more suitable as an HTL material.

The band edge alignment of an HTL material with that of the perovskite is an important parameter that can significantly affect the performance of the fabricated PVSC device. Figure \ref{electronic} presents the PBE-D3 computed vacuum-aligned band edges of the three DAHs compared to those of perovskites. MAPbI$_3$ and FAPbI$_3$ are computed to have similar band gaps of 1.6 eV at the PBE-D3 level, with the latter having a slightly larger ionization potential (IP) of 5.9 eV. Our PBE-D3 computed band gaps agree reasonably well with the reported literature values of 1.55 and 1.48 eV for MAPbI$_3$ and FAPbI$_3$, respectively.\cite{cs_13_2167} Moreover, our calculated higher IP of FAPbI$_3$ agrees with its experimentally observed higher oxidation stability than MAPbI$_3$.\cite{cs_13_2167} Similarly, DAT shows the highest IP of 4.5 eV. Considering the alignment of the band edges with both MAPbI$_3$ and FAPbI$_3$, all studied systems have VBMs that lie above those of the perovskites. This is desirable for the efficient hole migration from the perovskite layer to the HTL material after the photoexcitation. However, for DAT, the VBM aligns better with both MAPbI$_3$ and FAPbI$_3$ than DAP and DAS. Overall, based on our analyses for band edge alignments and IPs, DAT performs better as an HTL material than the other considered systems.

\subsubsection{3.2 Hole Mobility of the DAH Cores} \label{sec3.2}
\textbf{3.2 Hole Mobility of the DAH Cores}\newline
The semiclassical Marcus theory is widely used to calculate the carrier mobility in organic semiconductor materials \cite{zhang2015anisotropic, yuan2021solution, zhang2023shear}. Two important considerations in the Marcus theory are electronic coupling and reorganization energy. The electronic coupling refers to the intermolecular charge transport from one molecule to its near neighbors, while the reorganization energy ($\lambda$) refers to the energy penalty associated with geometry relaxations during charge transfer processes. The four-point scheme was used to calculate the reorganization energy of all systems.\cite{dell2022modeling} DAT is found to have the lowest $\lambda$ value of 0.3 eV compared to DAP (0.31 eV) and DAS (0.40 eV) (see Table \ref{table1}). For our systems with herringbone packing motifs, the electronic coupling depends not only on the relative intermolecular CoM distances but also on the orientations of the molecules in the dimers. Considering the calculated couplings between different dimers, DAT has the highest value of 0.21 eV for its near neighbor translational dimer (see SI Figure S7). On the other hand, the highest computed coupling values for DAP and DAS are 0.06 eV and 0.08 eV, respectively. Both $\lambda$ and the electronic coupling determine the overall calculated hole mobilities, although $\lambda$ carries more importance as it appears in the numerator of the decaying exponential (see the Methods section for more details). 

Using the semiclassical Marcus theory, the calculated hole mobility of DAT (0.73 cm$^2$/V$\cdot$s) was determined to be significantly higher than that of DAP (0.11 cm$^2$/V$\cdot$s) and DAS (0.16 {cm$^2$/V$\cdot$s}) cores (Table \ref{table1}). The enhanced hole mobility for DAT can be attributed to its ordered packing, strong $\pi$-$\pi$ interactions in the $b$ direction, and larger electronic couplings between its near neighbor dimers. Based on the superior properties of DAT as an HTL material, this material was selected for further structural modifications. As mentioned above, different functional groups and $\pi$-extensions were considered to decipher useful trends by probing their impacts on the structure and overall calculated electronic properties and hole mobilities.

\subsubsection{3.3 Hole Mobilities of the Modified DATs} \label{sec3.3}
\textbf{3.3 Hole Mobilities of the Modified DATs}\newline
\begin{figure*}[!h]
	\centering
	\includegraphics[width=0.99\linewidth]{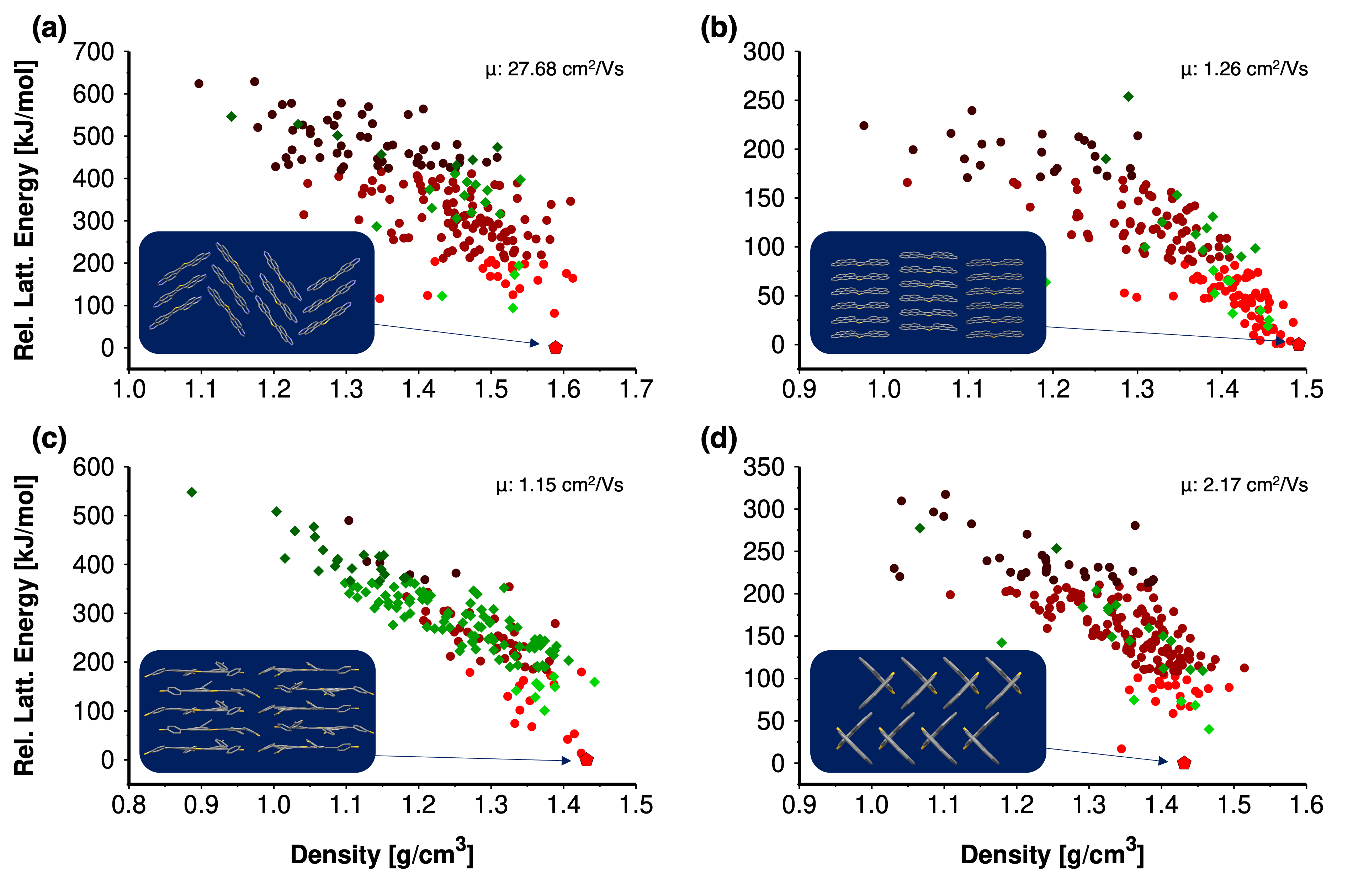}
	\caption{Calculated relative lattice energy vs. density correlation plots for (A) M7, (B) M1, (C) \textit{p}-thiophenyl substituted DAT, and (D) M6. PBE-D3 is used in all calculations except for M1, where PBE-MBD is employed. Crystal structures with and without translational dimer motifs are highlighted with red and green colors, respectively, with darker colors representing higher energy polymorphs.}
	\label{csp_corr_plot}
\end{figure*}
Significant efforts are devoted to finding novel organic molecular crystals with high hole mobilities for applications in electronics and optics.\cite{sokolov2011computational, nguyen2015density, nakata2023toward, nguyen2018remarkable} Here, an already high hole mobility OMC, i.e., DAT, was considered for further structural modifications. The performance of all functionalized systems is assessed here using two criteria: the ratio of the electronic coupling over $\lambda$ (V/$\lambda$) and the calculated hole mobilities ($\mu$) using the translational dimer model. All the distances between different dimers are the CoM distances. Naturally, the hole mobility will be higher for systems with higher electronic coupling but lower $\lambda$. Therefore, the aim here is to maximize electronic coupling values while minimizing $\lambda$. One should note that since the inner reorganization energy is considered only, the computed V/$\lambda$ ratios and mobilities constitute an overbound to the exact values. Nevertheless, since we are interested in identifying trends, the computed V/$\lambda$ values offer a valuable and effective screening tool for evaluating the considered systems. 

The calculated V/$\lambda$ ratios and $\mu$ for different functional group (FG) substituted DATs and ring-extended DATs are shown in the SI Figures S5 and S6, respectively. Overall, based on the screening metrics, all 74 considered FGs can be classified into four different categories: (i) small sulfur-containing FGs with the highest mobilities, (ii) moderately sized acyclic side groups with unsaturated bonds that give rise to good carrier transport, (iii) halogen-containing FGs exhibiting moderate mobilities, and (iv) bulky alkyl substituents, cyclic or acyclic, that result in poor mobilities. Typically, the electronic coupling exponentially decreases as the molecular separation increases. Hence, the smaller side groups, which do not put a large stress on the core structure, were found to perform better than heavier functional groups. In other words, bulky substituents are more likely to disrupt the inner strain of the DAT core and thus affect electron delocalization across the structure. Although larger in size, side groups containing double or triple bonds were found to produce large electronic couplings. The triple bond containing structures were also found to yield very low $\lambda$ values, which decrease with the increase in the number of the triple bonds, most likely due to their enhanced ability to accommodate the positive charge. Overall, functional groups with heavy atoms and unsaturated bonds demonstrated a lower bound of $\lambda$ values, whereas alkyl groups and esters were found to form an upper bound. On the other hand, the ring-extended structures and functional groups with aromatic rings demonstrated a higher value of V due to the $\pi$-extension.

\begin{figure}[!h]
	\centering
	\includegraphics[width=0.99\linewidth]{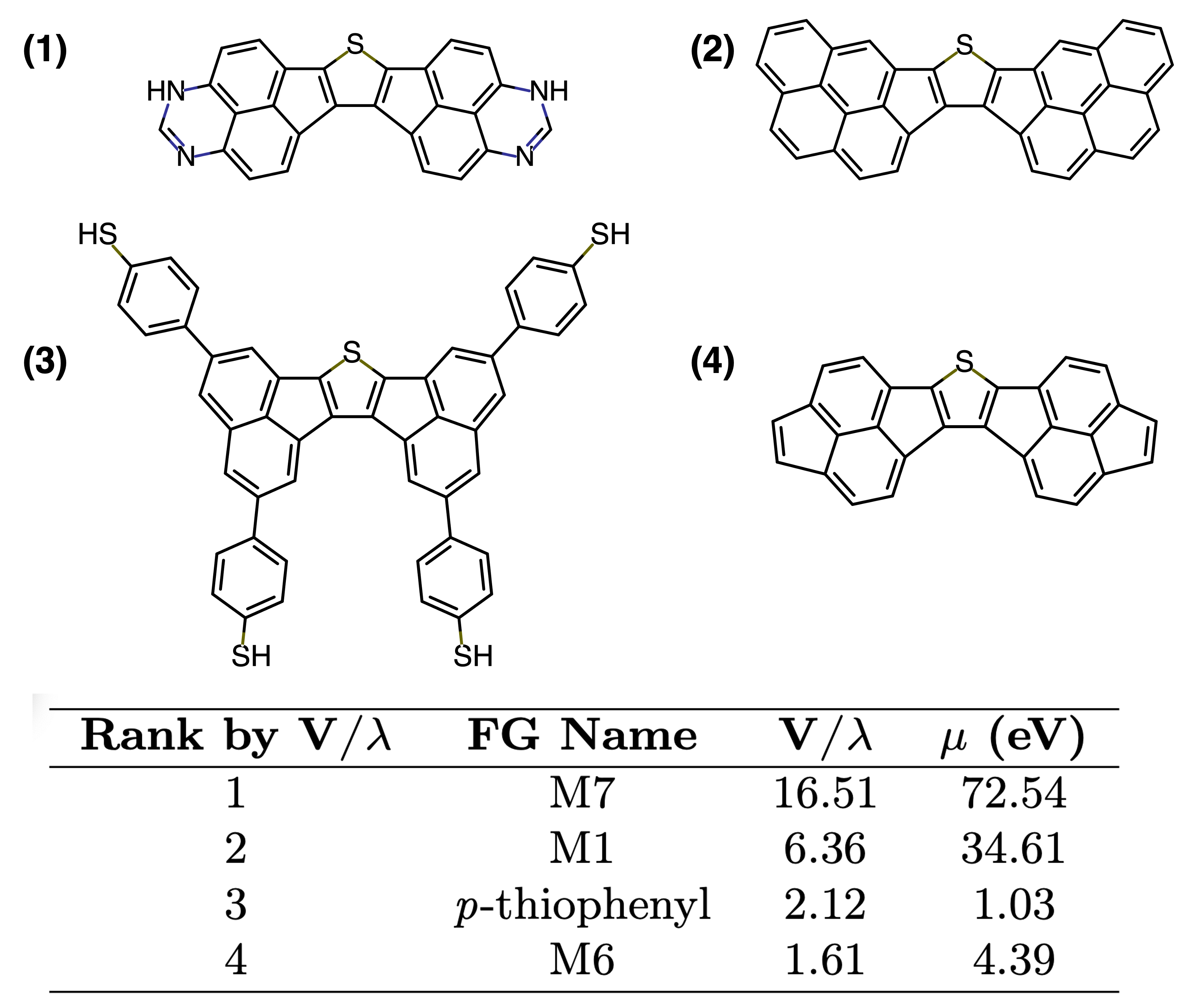}
	\caption{Top 4 promising candidates according to the V/$\lambda$ rankings obtained from applying the translational dimer model.}
	\label{top4}
\end{figure}

The top four best-performing FGs are shown in Figure \ref{top4}. 
Overall, three FGs were found to outperform the parent DAT system, including $p$-thiophenyl, thiomethyl, and thiol (see the SI Figure S8). Sulfur substituents outperformed other side groups considering both V/$\lambda$ ratios and hole mobilities. In both cases, three of the top four FGs contain sulfur. Moreover, when ranked by hole mobility, five out of the top ten FGs were found to contain sulfur atoms. Other than sulfur, chlorine and ether-containing FGs were also found to show high hole mobilities. For the ring-extended DAT systems, five out of ten structures outperformed the parent DAT system. The highest-performing structure (M7) and the second-highest (M8) are isomers with perimidine rings in their core. The third (M1) and fifth (M6) structures are $\pi$-extended with four extra 6-membered and two extra 5-membered fused aromatic rings. Overall, the effectiveness of $\pi$-extension through fused aromatic rings outperformed $\pi$-extension through functionalization (compare, for example, M7 to \textit{p}-thiophenyl substituted DAT). 
We note that the hole mobility of the functionalized DAT systems may be significantly influenced by thermal fluctuations, resulting in anisotropic transport and altered mobilities. To better understand this phenomenon, the calculation of the intermolecular vibrational modes is necessary for each system. However, that would require computationally intensive calculations, and it is therefore beyond the scope of the current study.

\subsubsection{3.4 Crystal Structure Predictions of the Top-Performing DATs} \label{sec3.3}
\textbf{3.4 Crystal Structure Predictions of the Top-Performing DATs}\newline
Once the V/$\lambda$ and $\mu$ for all functionalized and ring-extended DATs were screened, the structures were ranked to identify the ones with the highest hole mobility. Consequently, four representative systems, including the ring-extended DATs M7, M1, M6, and \textit{p}-thiophenyl substituted DAT were selected for crystal structure prediction based on their high V/$\lambda$ values (see Figure \ref{top4}). No solid-state structure has been reported experimentally for these systems, and no carrier mobilities have been reported for them either. As mentioned before, our CSP methodology was first validated by applying it to the parent DAT core with an experimentally reported crystal structure. Our methodology correctly predicted the experimental crystal structure of DAT as the lowest energy and highest density structure (see the SI Figure S10). The correlation plots for relative lattice energies vs. densities of all four top candidate HTLs are shown in Figure \ref{csp_corr_plot}. 
Interestingly, in all four of the lowest energy structures, a columnar packing with the translational dimer motif was found to be present. The lowest energy crystal structures and their calculated hole mobility are shown in the inset. It is worth mentioning that the relative lattice energy range is large in the lattice energy vs. density plot. This is because of the adopted workflow where all the candidates shown are the low-energy structures selected from their similarity clusters through AP clustering. As mentioned above, our own validation test for the parent DAT proves the accuracy of this approach for the considered family of OMCs; this method has also been used to successfully predict a number of other molecular crystals as well\cite{bier2021crystal, o2023ab}. In what follows, we provide more details on each of the four predicted crystal structures and their calculated electronic properties and carrier mobilities.

\begin{figure}
	\centering
	\includegraphics[width=0.99\linewidth]{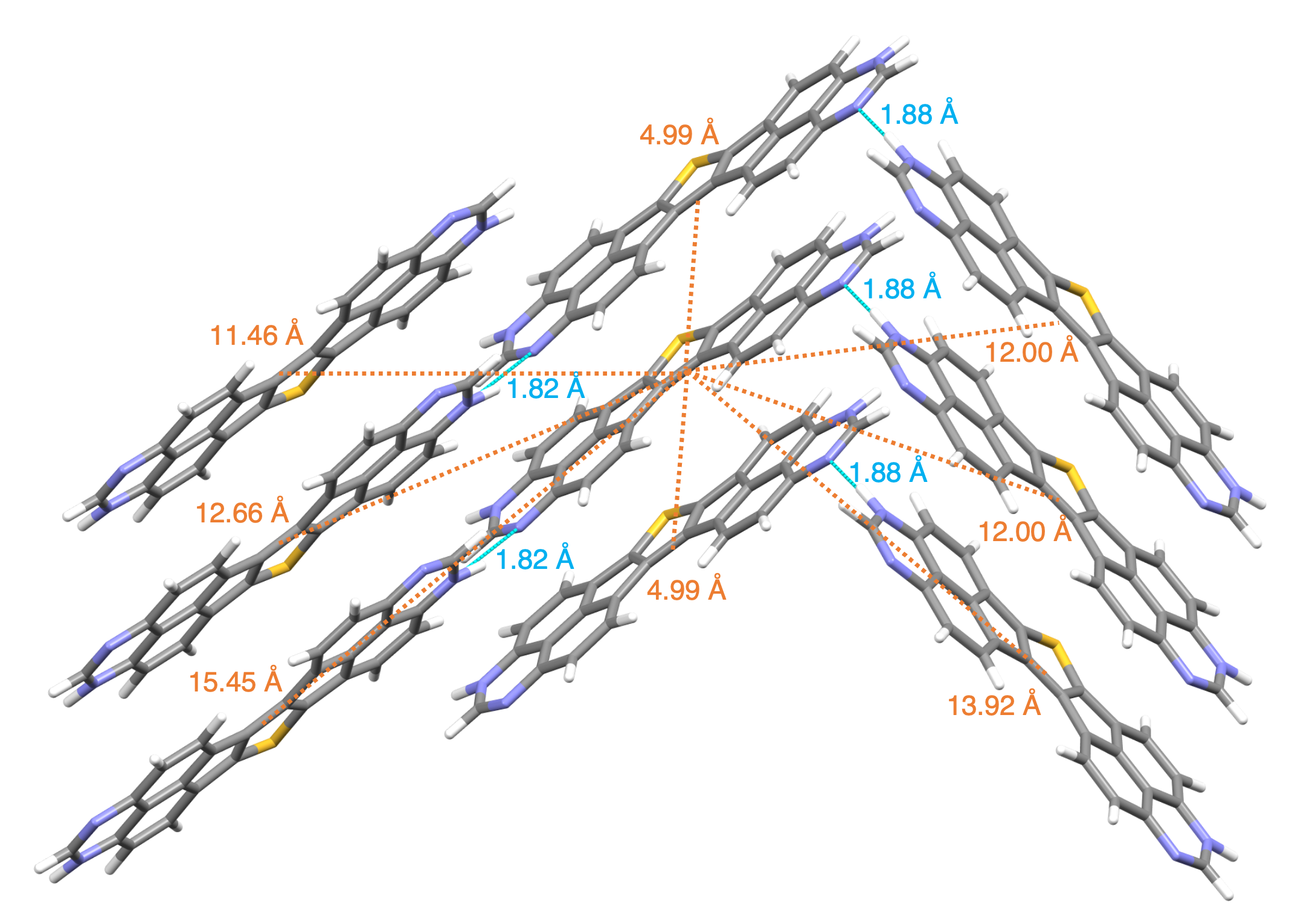}
	\caption{Calculated CoM distances (in orange) between a central molecule and its near neighbor units for the lowest energy crystal structure of M7. The calculated hydrogen bond distances (in blue) are also shown.}
	\label{h_bond}
\end{figure}

\textbf{M7}. As mentioned above, a columnar packing with the translational dimer motif is present in all lowest energy predicted crystal structures. For M7, the lowest energy structure shows significantly lower energy than the others (Figure \ref{csp_corr_plot}). Calculated hole mobility for the whole crystal structure using the semiclassical Marcus theory is 27.68 cm$^2$/V$\cdot$s while the predicted hole mobility using only the translational dimer model is 72.54 cm$^2$/V$\cdot$s for this system. The CoM distance along the translational dimer is 4.99~\AA~with an electronic coupling value of 0.02 eV (see Figure \ref{h_bond}). 
As one can intuitively imagine, hydrogen bonds should play a role in the packing of this structure. As shown in Figure \ref{h_bond}, hydrogen bonds are formed between the nearby columns. Indeed, because of the formation of these rather strong hydrogen bonds, the predicted structure has a density of 1.59 g/cm$^3$, which is the highest among all four predicted structures.
The building blocks of M7 are two perimidine structures. The core perimidine structure has already been synthesized.\cite{pozharskii2020perimidines, rewcastle2008pyrimidines} M7 can, therefore, be an ideal candidate for the synthesis, characterization, and fabrication of PVSCs with high power conversion efficiencies. 

\textbf{M1}. Similar to the other structures, the D3 dispersion correction was used for the final ranking of M1. However, these calculations resulted in multiple low-energy structures with very close energy differences (see SI Figure S11). To obtain a better resolution for our ranking, PBE-MBD calculations were performed for M1, the results of which are presented in Figure \ref{csp_corr_plot}b. The translational dimer motif is prevalent throughout the candidate structures, including in the predicted lowest energy one. This structure shows a sheetlike crystal packing. The calculated hole mobility for the predicted crystal structure is 1.26 cm$^2$/V$\cdot$s, with a 3.79~\AA~ CoM distance in the translational dimer direction. The CoM distance along the translational direction using the translation dimer model was 3.57~\AA. The predicted structure has the lowest energy and the highest density (1.49 g/cm$^3$, Figure \ref{csp_corr_plot}). The extended aromatic rings result in a closely packed $\pi$-stacking along the translational direction. The building block of M1, cyclopenta[cd]pyrene, is a well-known polyaromatic hydrocarbon that is already synthesized\cite{konieczny1979synthesis, murray1995quantitative} as well.

\textbf{\textit{p}-thiopenyl}. \textit{p}-thiophenyl substituted DAT has the largest molecular size among the studied four structures. Similar to the rest, the translational dimer motifs are present in the lowest energy structure, although overall, the structures without translational motifs are abundant in the high energy and low-density region (see Figure \ref{csp_corr_plot}(c)). Due to the four large side groups, the translational dimer motif was present at a much lower frequency. Calculated hole mobility for the whole crystal structure is 1.15 cm$^2$/V$\cdot$s, which is $>$50\% higher than the parent DAT (0.73 cm$^2$/V$\cdot$s). The CoM distance along the translational direction for the predicted structure is 4.75~\AA~ with a coupling value of 0.10 eV. Translational dimers are slipped to accommodate the bulky thiophenyl groups, forming a slipped sheetlike packing motif. The second lowest energy structure in the correlation plot has an energy difference of 13.9 kJ/mol from the lowest energy structure. It forms a sheetlike packing as well, but more slipped compared to the lowest energy structure (see SI Figure S12). As a result, it demonstrates a higher overlap and a higher coupling between near neighbors, resulting in a hole mobility value of 2.08 cm$^2$/V$\cdot$s. \textit{Tert-butyl} functionalized DAT with a similar substitution pattern as demonstrated in this work has been synthesized before \cite{li2018highly}. A similar synthetic approach can be adopted for synthesizing {\textit{p}-thiopenyl} substituted DAT.

\textbf{M6}. M6 is the smallest monomer of all four, containing two 5-membered aromatic rings fused to the DAT core structure. As shown in Figure \ref{csp_corr_plot}(d), the translational dimer motifs are prominent in most of the M6 crystal structures. As for the others, the lowest energy structure has a translational dimer motif with a full eclipse pattern and a hole mobility of 2.17 cm$^2$/V$\cdot$s. It forms a herringbone crystal packing with a CoM distance of 4.72~\AA~in between dimers in the translational direction. The reorganization energy for M6 is the highest (0.22 eV) among all four structures. However, the relatively large electronic coupling compensates for this. The building unit of M6, cyclopent[fg]acenaphthylene, has also been synthesized before\cite{Siegel_Tobe_Shinkai_2009}. Similar to the other three, this structure is also expected to be synthetically viable.
In addition to mobilities, the band edge alignments of all four predicted structures were calculated relative to MAPbI$_3$ and FAPbI$_3$ perovskites, where all showed good alignments similar to the DAT core (see SI Figure S13).


\section{4. Conclusions} \label{sec4}
\textbf{4. Conclusions}\newline
Of the 3 DAH structures, the unfunctionalized DAT core was found to exhibit the highest hole mobility. This correlated well with its calculated low hole effective mass, large VBM bandwidth, and low reorganization energy. Moreover, the VBM of DAT was found to align better with common perovskite materials such as MAPbI$_3$ and FAPbI$_3$. Using the translational dimer model, high-throughput screening calculations were performed for a total of 74 structurally modified DATs with different functional groups and $\pi$-extended systems. Those structures with higher hole mobilities than the parent DAT were selected for crystal structure predictions, which validated the accuracy of using the translational dimer model for high throughput screening.
As one of the limitations of our study, the semiclassical Marcus theory used here for describing the charge transfer mechanism neglects dynamic disorder originating from thermal fluctuations. In the hopping regime of the Marcus theory, the intermolecular electronic coupling is kept fixed. However, at room temperature, because of the weak van der Waals interactions in organic molecular crystals, the orientation of the molecule will fluctuate continuously as the charge transfer integral is modulated by nuclear motions. There are several approaches to include the dynamic disorder. For example, a phonon-assisted term can be added to the temperature dependence of mobility. Nevertheless, despite its limitations, the Marcus theory has been widely used to describe the charge transfer mechanism in various organic molecular crystals at room temperature and shows good agreement with experiments. 
This study provides useful insights into different factors contributing to high carrier mobilities in OMCs. In addition to extending the $\pi$-conjugation, the presence of heterocycles, functional groups such as sulfur atoms, and their orientations are also found to play crucial roles in carrier mobilities. Our results provide a clearer picture of the hole mobility mechanism in OMCs and offer useful insights for fast theoretical high-throughput screening and the inverse design of OMCs with high mobilities. Future works will involve the inclusion of dynamic disorder for further improvement of the calculated mobilities in OMCs.

\begin{acknowledgement}
This research was supported by the National Science Foundation through award no. DMR-2308895. Simulations used resources from Bridges-2\cite{brown2021bridges} at Pittsburgh Supercomputing Center through allocation CHE200007, CHE200008, and PHY230099 from the Advanced Cyberinfrastructure Coordination Ecosystem: Services \& Support (ACCESS) program,\cite{boerner2023access} which is supported by National Science Foundation grants \#2138259, \#2138286, \#2138307, \#2137603, and \#2138296. Technical support and computing resources provided by the HPC center at UMKC are also gratefully acknowledged.
\end{acknowledgement}

\begin{suppinfo}
Details of our calculations, including $k$-point paths for all DAHs, crystal structures of perovskites, band structures and projected density of states of DAHs, different functionalizations in DAT, V/$\lambda$ and hole mobility values of all 74 functionalized DATs, neighboring dimers for DAT, DAP, and DAS, benchmark of different XC functionals, a list of top-performing translational dimers for DAT, CSP validation (PBE-D3) for the parent DAT system, PBE-D3 calculated relative lattice energy vs. density correlation plots for M1, and PBE-D3 predicted second-ranked crystal structure of the \textit{p}-thiophenyl functionalized DAT. This material is available free of charge at pubs.acs.org.

\end{suppinfo}

\Addlcwords{is with of in the an a iv v as for on and by to spd-block}
\bibliography{bib.bib}

\end{document}


\newpage
\tableofcontents
\newpage

\section{Section S1. $k$-point paths of DAH cores  \label{cores}}
\addcontentsline{toc}{section}{Section S1. $k$-point paths of DAH cores}

\begin{figure*}[!htb]
\addcontentsline{toc}{figure}{Figure \ref{brillouin_zone}. $k$-point paths for DAT, DAP, and DAS}
\centering
\includegraphics[width=0.99\linewidth]{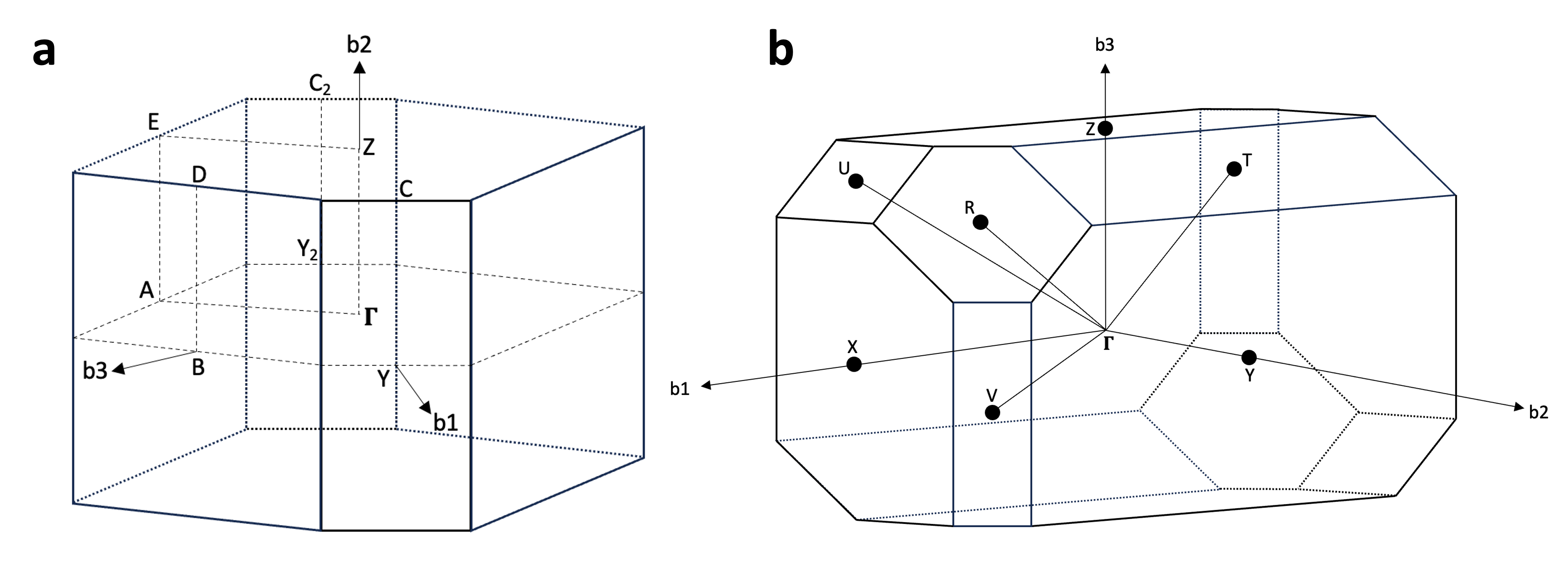}
\caption{$k$-point paths for (a) DAT, DAP, and (b) DAS.} 
\label{brillouin_zone}
\end{figure*}

\clearpage
\begin{figure*}[!htb]
\addcontentsline{toc}{figure}{Figure \ref{perovskites}. Crystal Structures of MAPbI$_3$ and FAPbI$_3$ perovskites}
\centering
\includegraphics[width=0.99\linewidth]{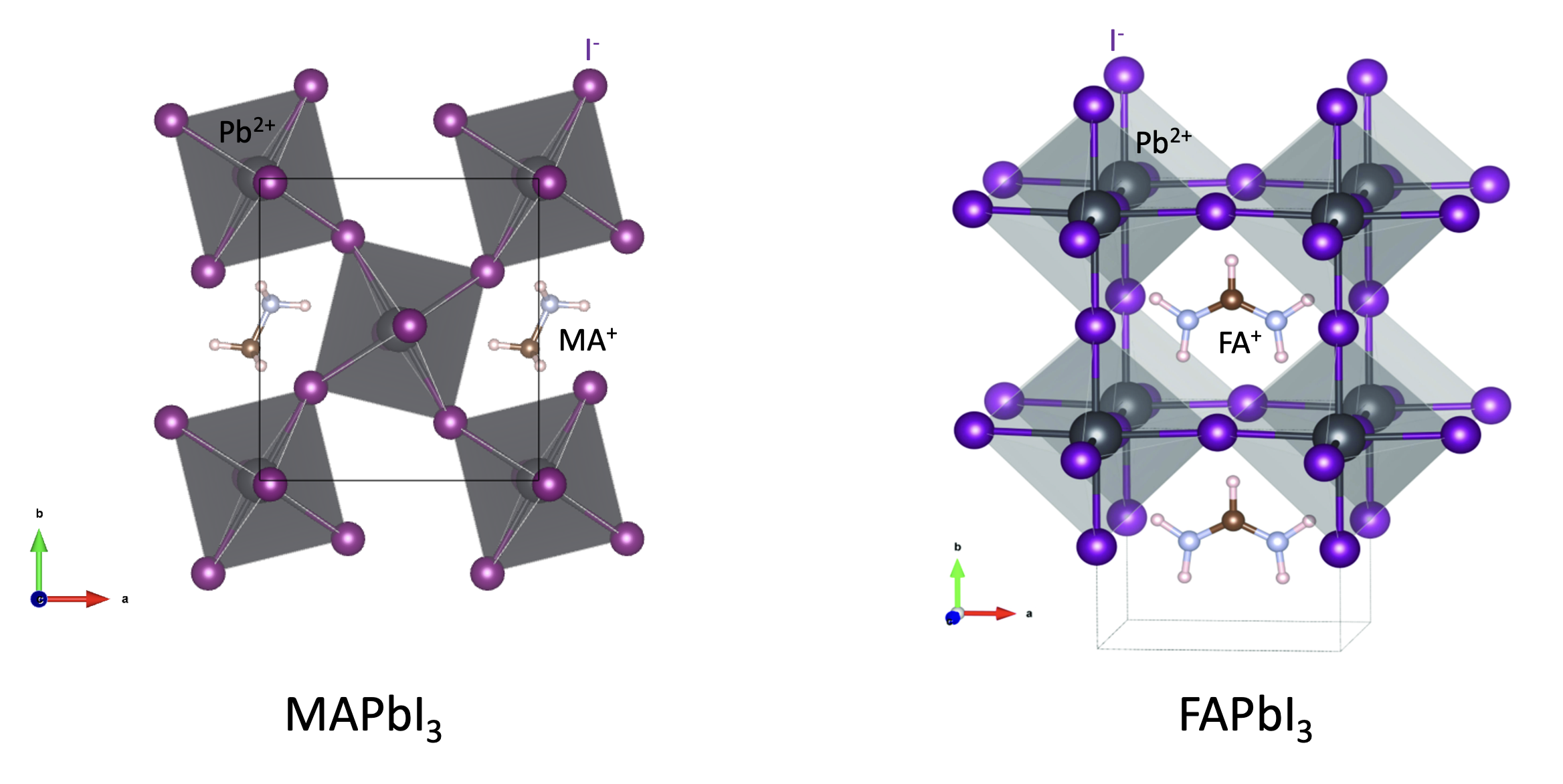}
\caption{Crystal Structures of MAPbI$_3$ and FAPbI$_3$ perovskites.} 
\label{perovskites}
\end{figure*}

\clearpage
\begin{figure*}[!htb]
\addcontentsline{toc}{figure}{Figure \ref{electronic}. Band Structure and Density of States for unfunctionalized DAT, DAP, and DAS}
\centering
\includegraphics[width=0.9\linewidth]{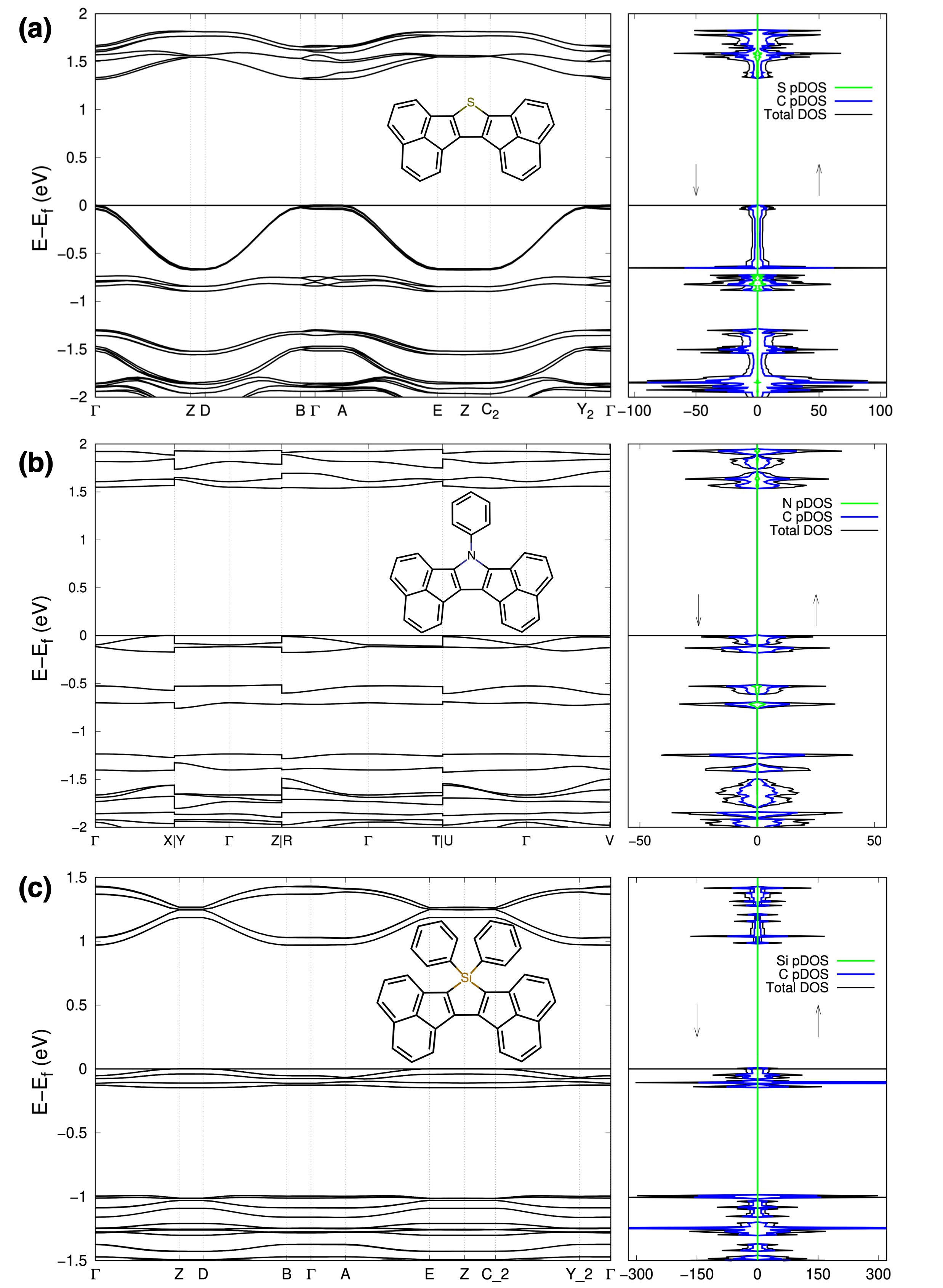}
\caption{PBE-D3 calculated band structures and projected density of states of (a) DAT, (b) DAP, and (c) DAS. The insets show the structure of monomers. The Fermi energy level is shown by horizontal lines at the zero position.} 
\label{electronic}
\end{figure*}

\clearpage
\section{Section S2. DAT Functionalizations \label{functionalization}}
\addcontentsline{toc}{section}{Section S2. DAT Functionalizations}

\begin{figure*}[!ht]
\addcontentsline{toc}{figure}{Figure \ref{functionalizing}. Functionalizing DAT core}
\centering
\includegraphics[width=0.7\linewidth]{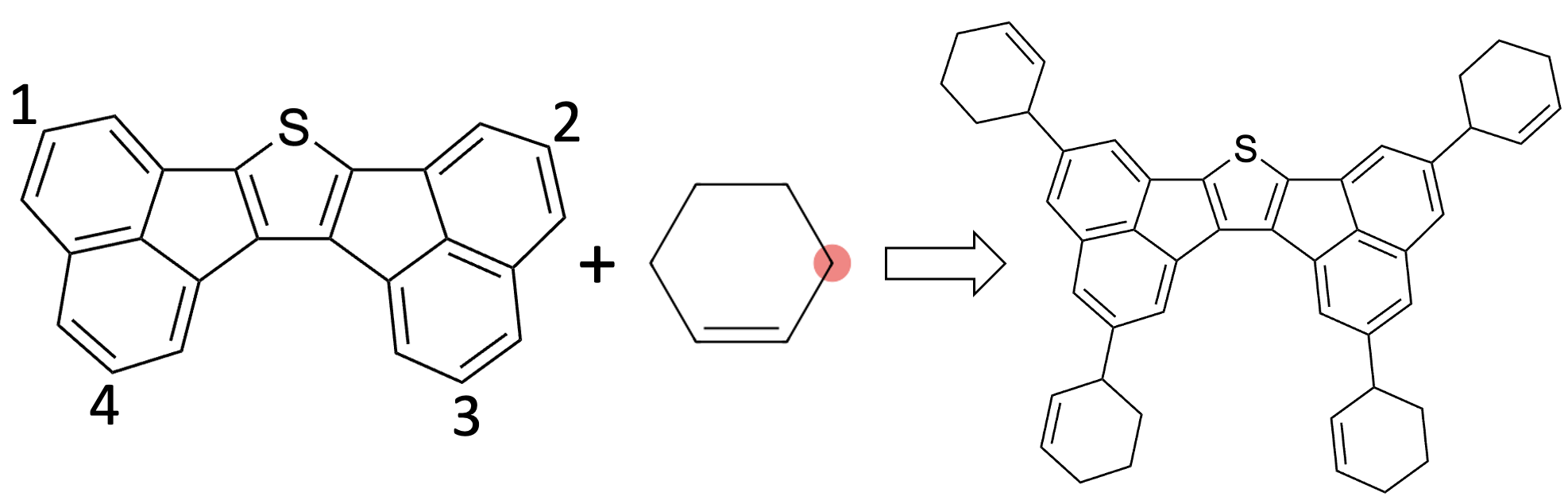}
\caption{DAT cores were functionalized in the numbered positions.} 
\label{functionalizing}
\end{figure*}

\clearpage

\clearpage
\begin{figure*}[!ht]
\centering
\includegraphics[width=0.9\linewidth]{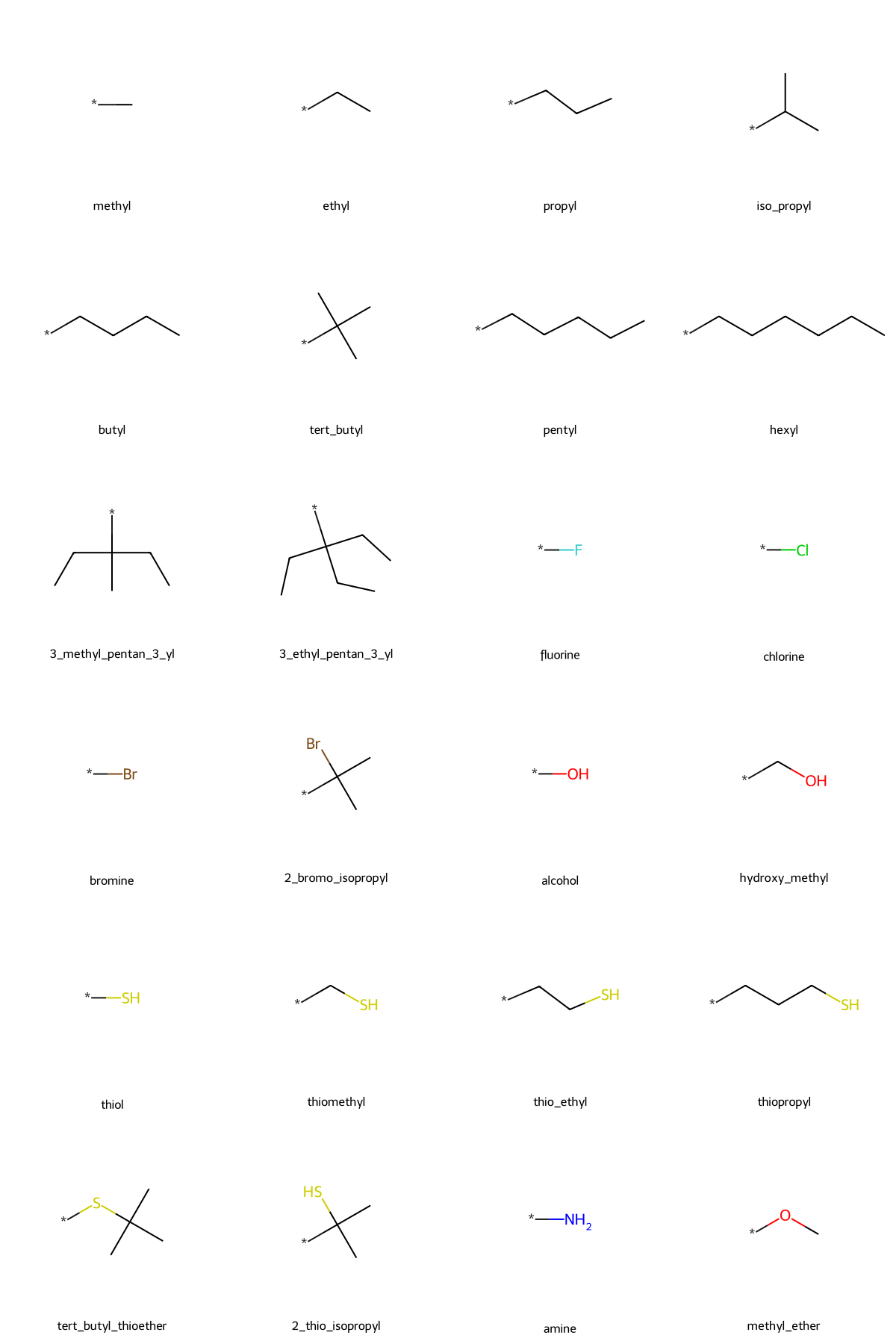}
\end{figure*}

\clearpage
\begin{figure*}[!ht]
\centering
\includegraphics[width=0.9\linewidth]{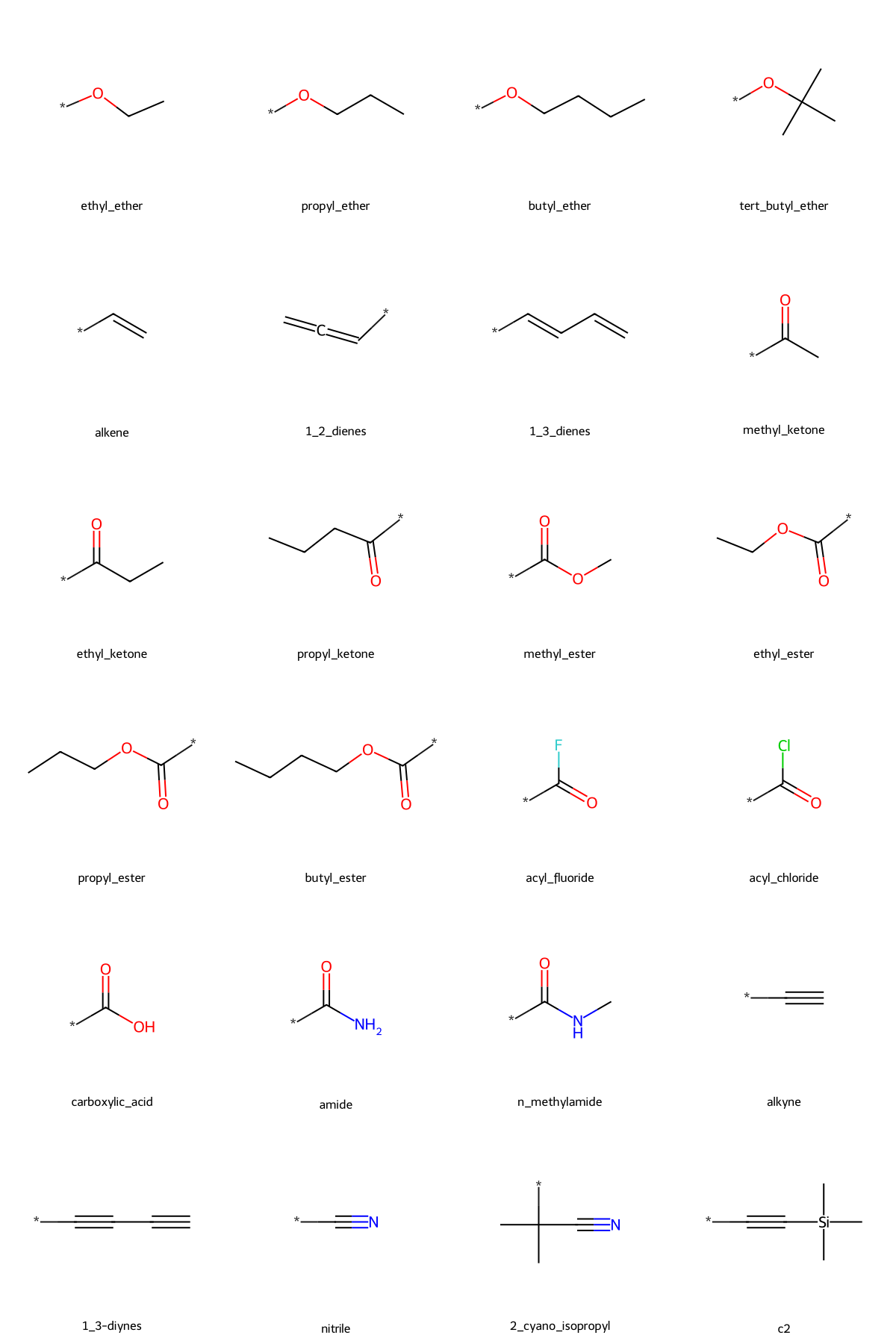}
\end{figure*}

\clearpage
\begin{figure*}[!ht]
\addcontentsline{toc}{figure}{Figure \ref{all_fg}. Comprehensive list of all functional groups}
\centering
\includegraphics[width=0.9\linewidth]{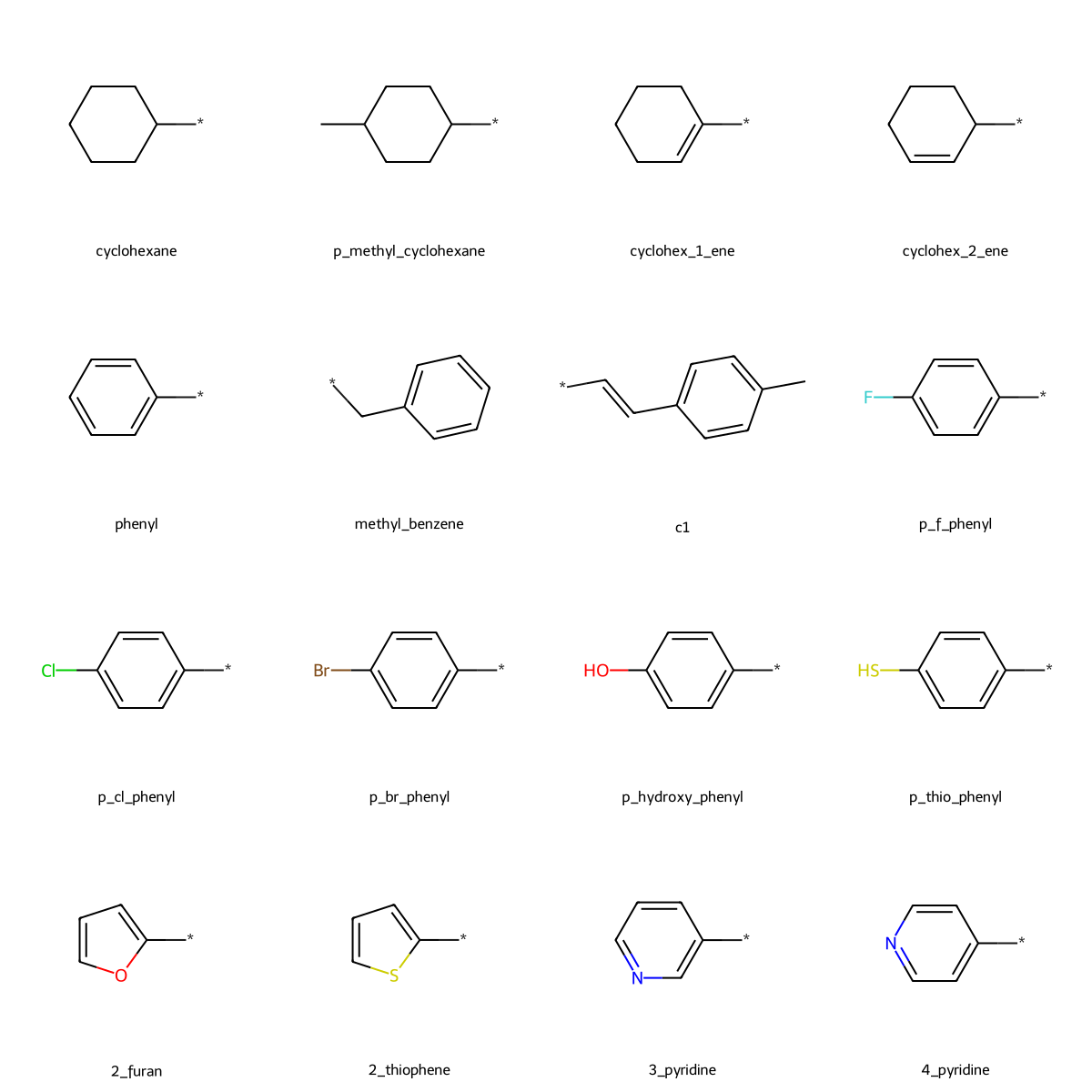}
\caption{List of all functional groups. * represents the connecting point with the DAT core.} 
\label{all_fg}
\end{figure*}

\clearpage
\begin{figure*}[!ht]
\addcontentsline{toc}{figure}{Figure \ref{dat_deriv}. DAT derivatives generated by aromatic ring extension.}
\centering
\includegraphics[width=0.99\linewidth]{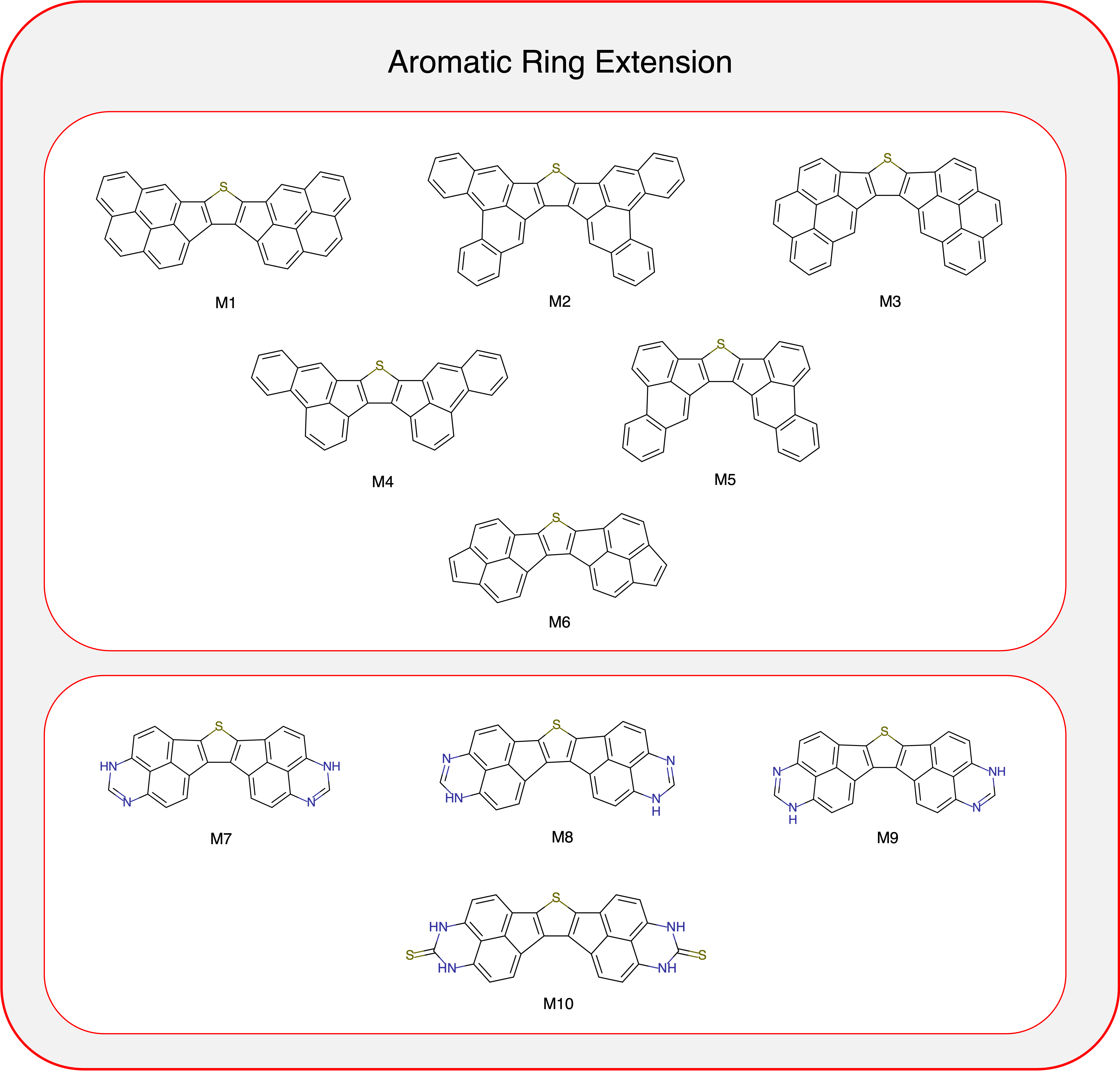}
\caption{DAT derivatives generated by aromatic ring extension. The first group of structures contains no heteroatoms, while the second group contains N in the aromatic ring.} 
\label{dat_deriv}
\end{figure*}

\clearpage
\begin{table}
\centering
\addcontentsline{toc}{table}{Table \ref{all_fg_values}. V/$\lambda$ and hole mobility values for all 64 functionalized DAT systems}
\centering
    \resizebox{0.99\textwidth}{!}{
    \begin{tabular}{cccc|cccc} \hline  
Sl & FG & V/$\lambda$ & $\mu$ (${cm^2/Vs}$) & Sl & FG & V/$\lambda$ & $\mu$ (${cm^2/Vs}$) \\ \hline
1 & methyl & 1.6E-01 & 5.7E-04 & 33 & ethyl\_ketone & 3.2E-02 & 1.8E-05 \\
2 & ethyl & 3.6E-02 & 4.0E-05 & 34 & propyl\_ketone & 2.1E-02 & 9.0E-05 \\
3 & propyl & 1.3E-02 & 8.0E-06 & 35 & methyl\_ester & 7.6E-02 & 5.9E-05 \\
4 & iso\_propyl & 6.7E-02 & 4.6E-04 & 36 & ethyl\_ester & 4.9E-02 & 2.4E-05 \\
5 & butyl & 2.7E-02 & 5.6E-05 & 37 & propyl\_ester & 4.0E-03 & 5.9E-07 \\
6 & tert\_butyl & 3.5E-02 & 6.9E-02 & 38 & butyl\_ester & 3.3E-02 & 4.3E-05 \\
7 & pentyl & 2.3E-02 & 1.1E-04 & 39 & acyl\_fluoride & 1.5E-01 & 3.6E-04 \\
8 & hexyl & 6.7E-02 & 2.4E-04 & 40 & acyl\_chloride & 1.4E-01 & 1.7E-04 \\
9 & 3\_methyl\_pentan\_3\_yl & 5.8E-02 & 4.3E-04 & 41 & carboxylic\_acid & 1.5E-01 & 2.8E-04 \\
10 & 3\_ethyl\_pentan\_3\_yl & 1.4E-02 & 1.7E-05 & 42 & amide & 1.4E-01 & 2.1E-04 \\
11 & fluorine & 2.0E-01 & 7.9E-04 & 43 & n\_methylamide & 8.8E-02 & 1.2E-04 \\
12 & chlorine & 6.1E-01 & 4.9E-01 & 44 & alkyne & 2.0E-01 & 1.3E-03 \\
13 & bromine & 2.0E-01 & 1.0E-03 & 45 & 1\_3-diynes & 2.0E-01 & 3.0E-03 \\
14 & 2\_bromo\_isopropyl & 3.7E-02 & 3.4E-03 & 46 & nitrile & 1.8E-01 & 6.6E-04 \\
15 & alcohol & 2.2E-01 & 1.5E-03 & 47 & 2\_cyano\_isopropyl & 1.7E-02 & 7.9E-06 \\
16 & hydroxy\_methyl & 1.3E-01 & 2.5E-04 & 48 & c2 & 5.5E-02 & 1.6E-04 \\
17 & thiol & 7.2E-01 & 8.5E-01 & 49 & cyclohexane & 2.7E-02 & 1.2E-04 \\
18 & thiomethyl & 8.4E-01 & 1.0E+00 & 50 & p\_methyl\_cyclohexane & 1.3E-02 & 1.1E-05 \\
19 & thio\_ethyl & 1.1E-01 & 6.7E-02 & 51 & cyclohex\_1\_ene & 7.1E-02 & 3.3E-04 \\
20 & thiopropyl & 2.8E-02 & 4.7E-03 & 52 & cyclohex\_2\_ene & 2.3E-02 & 2.2E-05 \\
21 & tert\_butyl\_thioether & 1.7E-01 & 1.0E-01 & 53 & phenyl & 2.9E-01 & 4.0E-03 \\
22 & 2\_thio\_isopropyl & 8.8E-02 & 4.9E-02 & 54 & methyl\_benzene & 1.5E-02 & 2.5E-05 \\
23 & amine & 2.2E-01 & 1.9E-03 & 55 & c1 & 1.6E-01 & 1.4E-02 \\
24 & methyl\_ether & 1.4E-01 & 3.2E-04 & 56 & p\_f\_phenyl & 1.6E-01 & 1.5E-03 \\
25 & ethyl\_ether & 1.1E-01 & 1.0E-03 & 57 & p\_cl\_phenyl & 4.9E-01 & 5.5E-01 \\
26 & propyl\_ether & 5.3E-02 & 3.4E-04 & 58 & p\_br\_phenyl & 1.8E-01 & 2.1E-03 \\
27 & butyl\_ether & 2.0E-02 & 1.7E-04 & 59 & p\_hydroxy\_phenyl & 1.6E-01 & 3.3E-03 \\
28 & tert\_butyl\_ether & 3.0E-01 & 3.0E-01 & 60 & p\_thio\_phenyl & 1.0E+00 & 2.1E+00 \\
29 & alkene & 1.9E-01 & 2.6E-03 & 61 & 2\_furan & 2.2E-01 & 1.3E-02 \\
30 & 1\_2\_dienes & 1.7E-01 & 2.1E-03 & 62 & 2\_thiophene & 2.0E-01 & 5.1E-03 \\
31 & 1\_3\_dienes & 1.8E-01 & 1.5E-03 & 63 & 3\_pyridine & 1.4E-01 & 1.2E-03 \\
32 & methyl\_ketone & 8.5E-02 & 9.9E-05 & 64 & 4\_pyridine & 1.6E-01 & 7.8E-04 \\ \hline
\end{tabular}}
    \caption{V/$\lambda$ and hole mobility values for all 64 functionalized DAT systems.}
    \label{all_fg_values}
\end{table}

\clearpage
\begin{table}
\centering
\addcontentsline{toc}{table}{Table \ref{ring_ext_values}. V/$\lambda$ and hole mobility values for all ring extended DAT systems}
\centering
\resizebox{0.4\textwidth}{!}{
\begin{tabular}{cccc} \hline  
Sl & Structure & V/$\lambda$ & $\mu$ (${cm^2/Vs}$) \\ \hline
1 & M1 & 6.4E+00 & 3.5E+01 \\
2 & M2 & 3.1E-01 & 3.6E-03 \\
3 & M3 & 5.7E-01 & 1.5E-03 \\
4 & M4 & 6.9E-02 & 8.7E-03 \\
5 & M5 & 5.5E-01 & 6.6E-03 \\
6 & M6 & 1.6E+00 & 4.4E+00 \\
7 & M7 & 1.7E+01 & 7.3E+01 \\
8 & M8 & 1.1E+01 & 5.7E+01 \\
9 & M9 & 4.4E-01 & 3.0E-02 \\
10 & M10 & 1.7E+00 & 2.6E+00 \\ \hline
\end{tabular}}
\caption{V/$\lambda$ and hole mobility values for all ring extended DAT systems.}
\label{ring_ext_values}
\end{table}

\clearpage
\begin{figure*}[!ht]
\addcontentsline{toc}{figure}{Figure \ref{figS_dist_all_dah}. All near neighbor dimers in DAT, DAP, and DAS}
\centering
\includegraphics[width=0.99\linewidth]{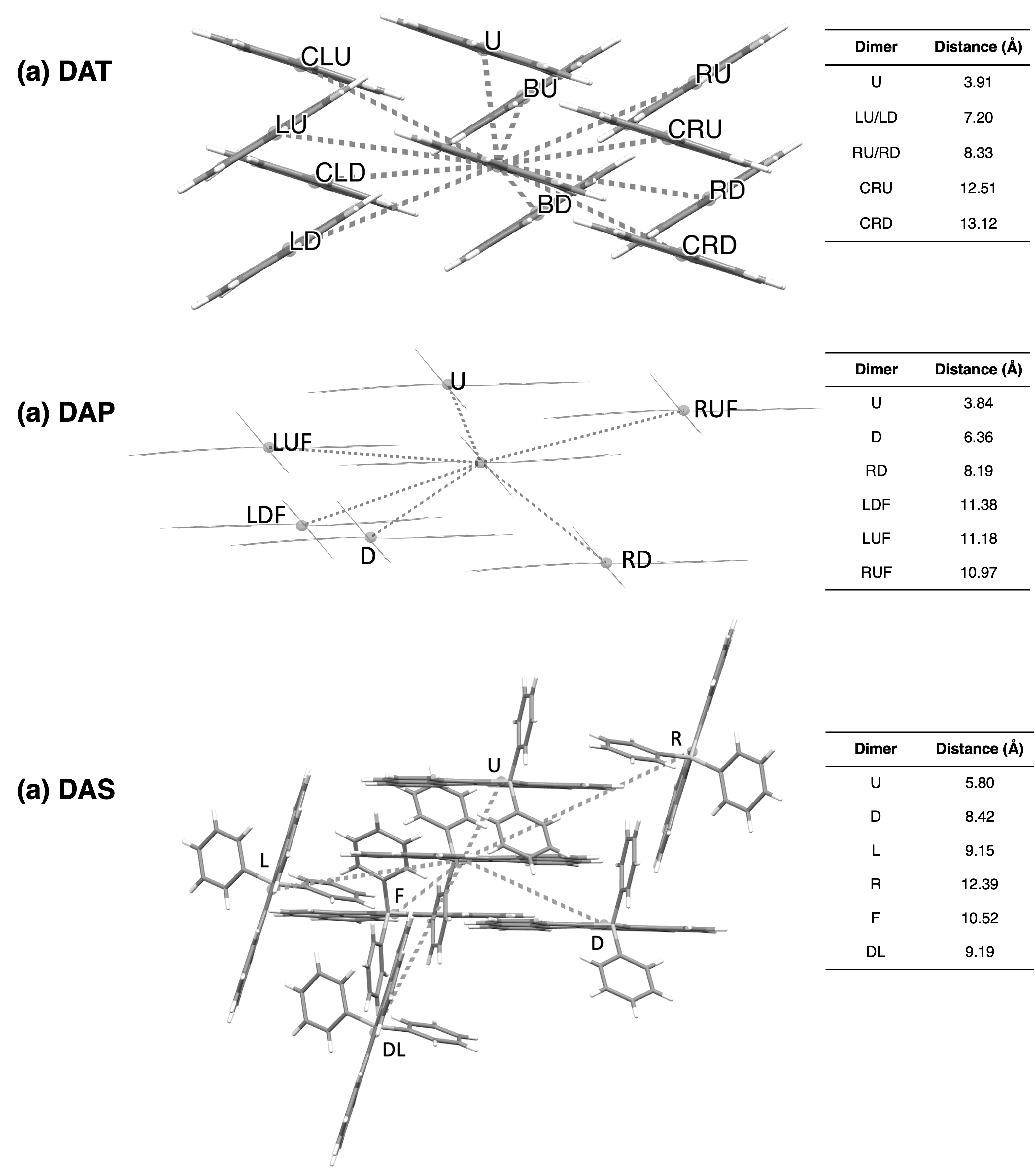}
\caption{All near neighbor dimers in (a) DAT, (b) DAP, and (c) DAS experimental crystal structures.} 
\label{figS_dist_all_dah}
\end{figure*}

\clearpage
\begin{table}
\addcontentsline{toc}{table}{Table \ref{tableSdatFunctionals}. Performance of different XC functionals for DAT, DAP, and DAT}
\centering
    \resizebox{0.99\textwidth}{!}{
    \begin{tabular}{cccccc} \hline
    System & Method & Cell Length (a)& Cell Length (b) & Cell Length (c) & Nearest neighbor distance (\AA) \\ \hline 
    DAT & PBE & 2.7 & 8.1 & 1.8 & 8.1 \\ 
    & RPBE & 7.2 & 11.3 & 4.2 & 11.3 \\ 
    & PBEsol & 1.3 & 2.3 & 1.0 & 2.3 \\ 
    & PBE-D2 & -2.1 & -5.6 & -1.8 & -5.6 \\ 
    & PBE-D3(Zero Damping)& -0.8 & 0.4 & -0.8 & 0.4 \\  
    & PBE-D3(BJ Damping) & -0.8 & -2.1 & -0.6 & -2.1 \\ 
    & PBE-TS & -0.8 & 0.4 & -0.8 & 0.4 \\
    & PBE-MBD & -0.8 & 0.4 & -0.8 & 0.4 \\ \hline
    DAP & PBE & 5.4 & 2.1 & 5.2 & 7.4 \\ 
    & RPBE & 9.3 & 5.1 & 8.1 & 11.8 \\ 
    & PBEsol & 1.4 & 0.8 & 1.5 & 2.1 \\ 
    & PBE-D2 & -5.8 & -2.0 & -2.4 & -4.7 \\ 
    & PBE-D3(Zero Damping) & -0.8 & -0.0 & 0.3 & 0.3 \\ 
    & PBE-D3(BJ Damping) & -2.1 & -0.4 & -0.6 & -1.3 \\ 
    & PBE-TS & -2.8 & -0.5 & -0.8 & -3.3 \\
    & PBE-MBD & -1.5 & -0.3 & -0.3 & -2.0 \\ \hline
    DAS & PBE & 3.15 & 5.09 & 3.09 & 4.64 \\ 
    & RPBE & 7.13 & 8.86 & 5.30 & 9.31 \\  
    & PBEsol & 0.75 & 0.91 & 1.38 & 1.71 \\  
    & PBE-D2 & -1.98 & -3.89 & -1.80 & -2.21 \\  
    & PBE-D3 (Zero Damping) & -0.35 & -0.41 & -0.17 & 0.14 \\  
    & PBE-D3 (BJ Damping) & -0.43 & -1.78 & -0.35 & -0.38 \\ 
    & PBE-TS & -1.0 & -1.7 & -1.2 & -1.9 \\
    & PBE-MBD & -0.3 & -0.4 & -0.2 & -0.1 \\ \hline
    \end{tabular}}
    \caption{Comparisons among the performance of different XC functionals in terms of the \% change with respect to the experimental crystal structures of DAT, DAP, and DAS.}
    \label{tableSdatFunctionals}
\end{table}

\clearpage
\section{Section S3. DAT Dimer Calculations \label{secS3}}
\addcontentsline{toc}{section}{Section S3. DAT Dimer Calculations}

\begin{table}
\centering
\resizebox{0.99\textwidth}{!}{
\begin{tabular}{|c|ccc|ccc|ccc|ccc|} \hline
\multicolumn{4}{|c|}{GFN1-xTB} & \multicolumn{3}{|c|}{PBE-D3 Zero Damping} & \multicolumn{3}{|c|}{PBE-TS} & \multicolumn{3}{|c|}{PBE-MBD} \\ \hline
Sl & Dist & Angle & Rel Energy (kJ/mol) & Dist & Angle & Rel Energy (kJ/mol) & Dist & Angle & Rel Energy (kJ/mol) & Dist & Angle & Rel Energy (kJ/mol) \\ \hline
1 & 3.4 & 180 & -16.5 & 3.7 & 180 & -11.3 & 3.6 & 0 & -39.6 & 3.6 & 180 & -14.5 \\
2 & 3.3 & 180 & -16 & 3.6 & 180 & -10.5 & 3.5 & 180 & -27 & 3.7 & 180 & -13.6 \\
3 & 3.5 & 180 & -14.9 & 3.8 & 180 & -10 & 3.6 & 180 & -27 & 3.5 & 180 & -13 \\
4 & 3.2 & 180 & -12.4 & 3.9 & 180 & -7.6 & 3.4 & 180 & -25.5 & 3.8 & 180 & -10.9 \\
5 & 3.6 & 180 & -12 & 3.5 & 180 & -6.8 & 3.7 & 180 & -21 & 3.4 & 180 & -8.2 \\
6 & 3.5 & 0 & -10.9 & 4 & 180 & -4.3 & 3.3 & 180 & -17 & 3.9 & 180 & -7.4 \\
7 & 3.4 & 0 & -9.9 & 4.1 & 180 & -0.3 & 3.8 & 180 & -13.7 & 4 & 180 & -3.4 \\
8 & 3.6 & 0 & -9.7 & \textbf{3.9} & \textbf{0} & \textbf{0} & 3.7 & 0 & -7.7 & 3.8 & 0 & -0.5 \\
9 & 3.7 & 180 & -8.4 & 3.4 & 180 & 0.4 & 3.9 & 180 & -7 & \textbf{3.9} & \textbf{0} & \textbf{0} \\
10 & 3.7 & 0 & -7.1 & 3.8 & 0 & 0.5 & 3.8 & 0 & -3.9 & 3.7 & 0 & 0.2 \\
11 & 3.3 & 0 & -5.6 & 4 & 0 & 0.7 & 3.5 & 0 & -1.1 & 4.1 & 180 & 1.2 \\
12 & 3.8 & 180 & -4.4 & 3.7 & 0 & 2.7 & 4 & 180 & -0.7 & 3.3 & 180 & 1.2 \\
13 & 3.1 & 180 & -4.1 & 4.1 & 0 & 3.3 & \textbf{3.9} & \textbf{0} & \textbf{0} & 4 & 0 & 1.5 \\
14 & 3.8 & 0 & -3.8 & 3.6 & 0 & 8.4 & 4 & 0 & 0.9 & 3.6 & 0 & 4.3 \\
15 & 3.9 & 180 & -0.2 & 3.3 & 180 & 12.7 & 3.2 & 180 & 2.8 & 4.1 & 0 & 4.7 \\
16 & \textbf{3.9} & \textbf{0} & \textbf{0} & 3.5 & 0 & 19.6 & 4.1 & 180 & 6 & 3.5 & 0 & 13.4 \\
17 & 3.2 & 0 & 4 & 3.2 & 180 & 33 & 4.1 & 0 & 8.9 & 3.2 & 180 & 18 \\
18 & 4 & 0 & 4 & 3.4 & 0 & 36.6 & 3.4 & 0 & 12.5 & 3.4 & 0 & 27.9 \\
19 & 4 & 180 & 4.1 & 3.3 & 0 & 61.8 & 3.3 & 0 & 33.7 & 3.1 & 180 & 44.6 \\
20 & 4.1 & 0 & 8.1 & 3.1 & 180 & 64.5 & 3.1 & 180 & 34.7 & 3.3 & 0 & 50.1 \\ \hline
\end{tabular}}
\caption{Comparison among the performance of GFN1-xTB, PBE-D3, PBE-TS, and PBE-MBD for ranking the relative energies (in kJ/mol) of the top 20 lowest energy DAT dimers. The bold-faced values represent the dimer distances and angles of the experimental structure. The energy of the experimental dimer is set to 0.}
\label{xtb_compare}
\addcontentsline{toc}{table}{Table \ref{xtb_compare}. Performance of GFN1-xTB vs. PBE-D3, PBE-TS, and PBE-MBD}
\end{table}

\clearpage
\section{Section S4. Crystal Structure Predictions \label{secS4}}
\addcontentsline{toc}{section}{Section S4. Crystal Structure Predictions}

\begin{figure*}[!ht]
\addcontentsline{toc}{figure}{Figure \ref{dat_csp}. CSP for the parent DAT system.}
\centering
\includegraphics[width=0.99\linewidth]{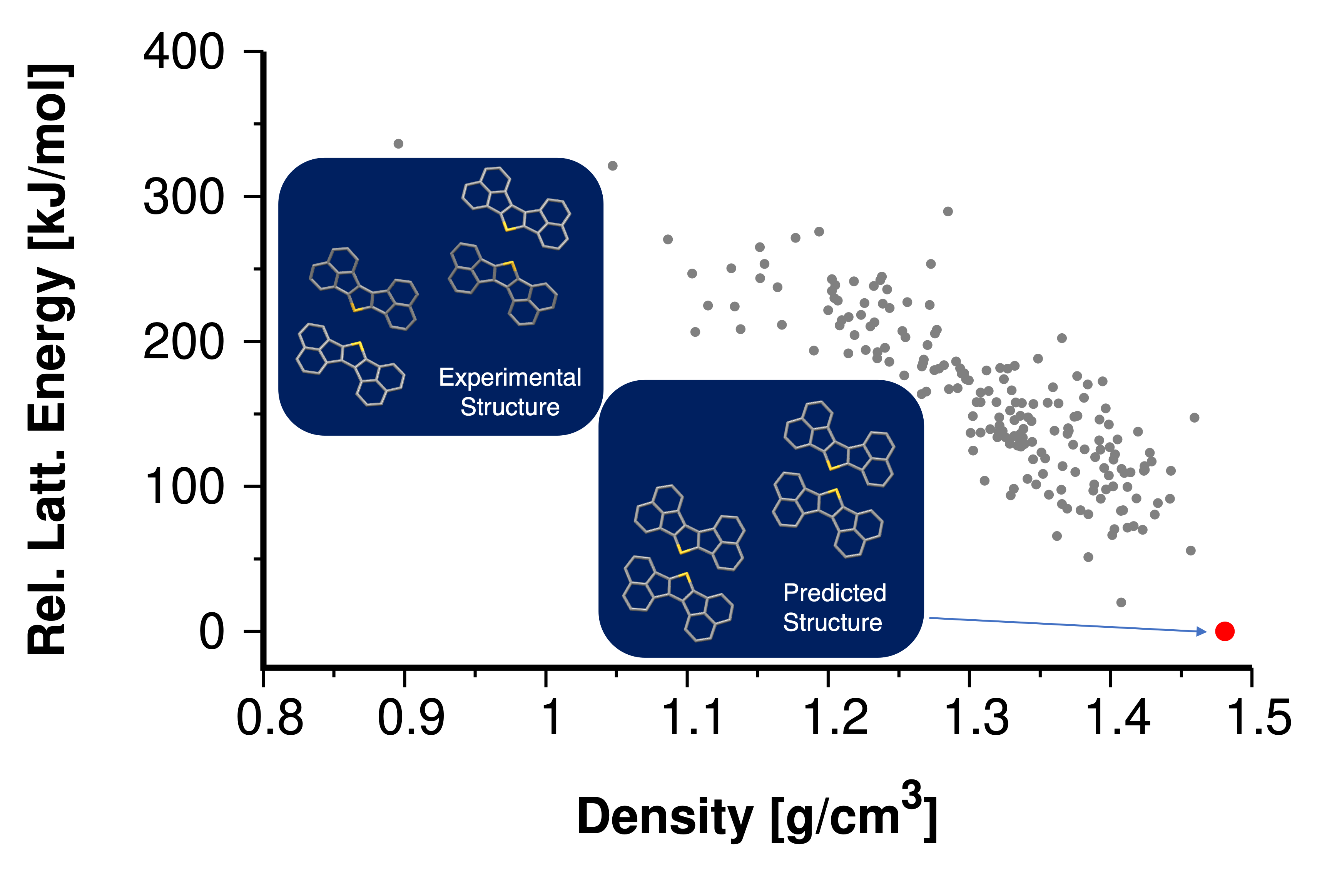}
\caption{CSP validation (PBE-D3) for the parent DAT system. The predicted lowest energy crystal structure is shown in red.} 
\label{dat_csp}
\end{figure*}

\clearpage
\begin{figure*}[!ht]
\addcontentsline{toc}{figure}{Figure \ref{p_thio_rank2}. PBE-D3 calculated relative lattice energy vs. density for M1.}
\centering
\includegraphics[width=0.7\linewidth]{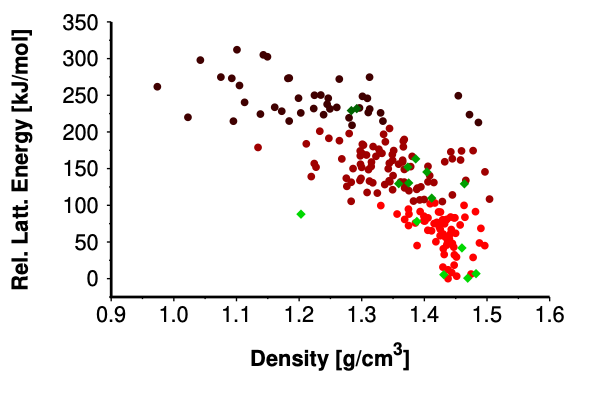}
\caption{PBE-D3 calculated relative lattice energy vs. density correlation plots for M1. Crystal structures with and without translational dimer motifs are highlighted with red and green colors, respectively, with darker colors representing higher energy polymorphs.} 
\label{p_thio_rank2}
\end{figure*}

\clearpage
\begin{figure*}[!ht]
\addcontentsline{toc}{figure}{Figure \ref{p_thio_rank2}. Predicted second-ranked crystal structure of the \textit{p}-thiophenyl functionalized DAT.}
\centering
\includegraphics[width=0.7\linewidth]{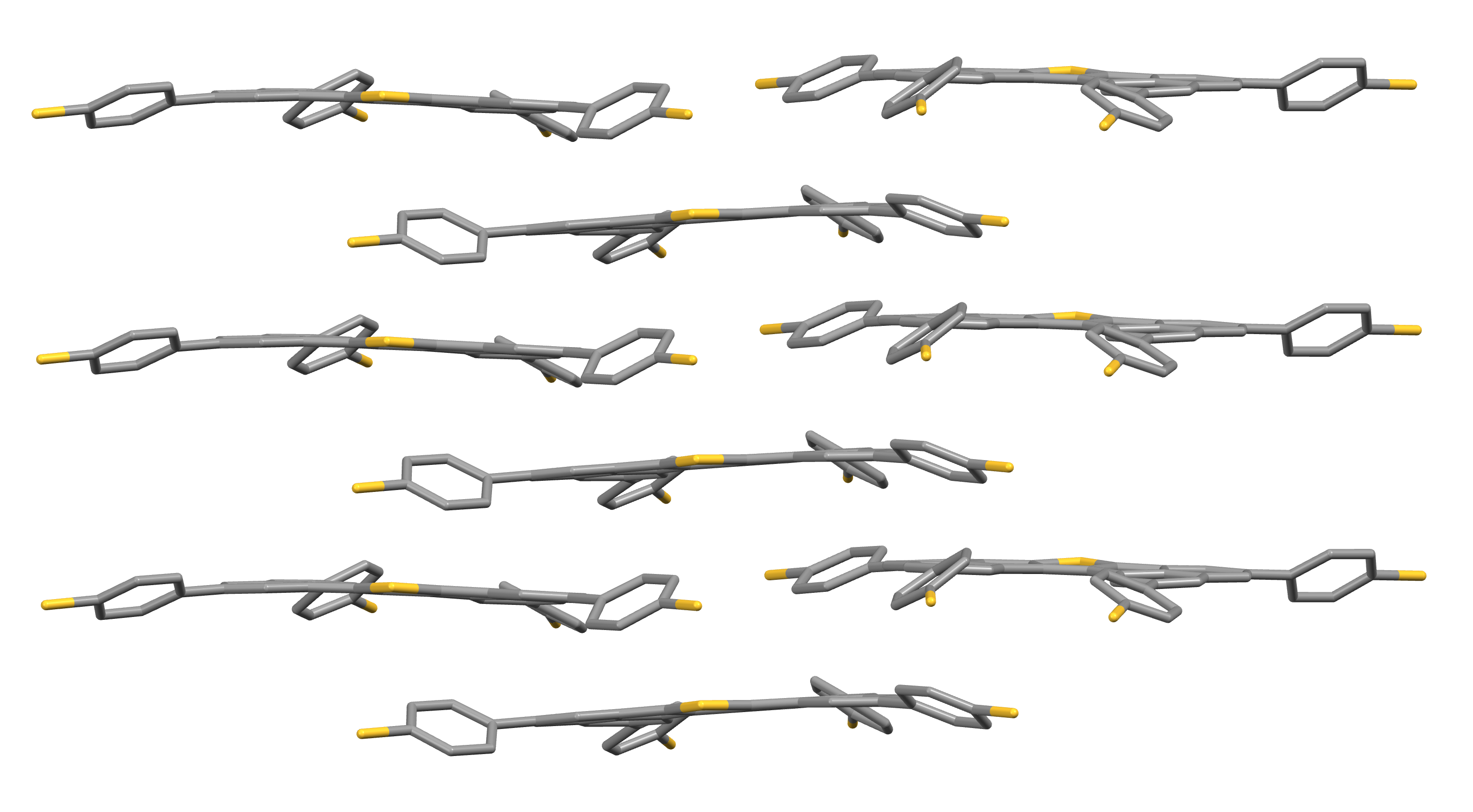}
\caption{PBE-D3 predicted second-ranked crystal structure of the \textit{p}-thiophenyl functionalized DAT.} 
\label{p_thio_rank2}
\end{figure*}

\clearpage
\begin{figure*}[!h]
\addcontentsline{toc}{figure}{Figure \ref{electronic_top}. PBE-D3 calculated vacuum-aligned band edges and band gaps for the top functionalized candidates compared to MAPbI$_3$ and FAPbI$_3$ perovskites.}
	\centering
	\includegraphics[width=0.9\linewidth]{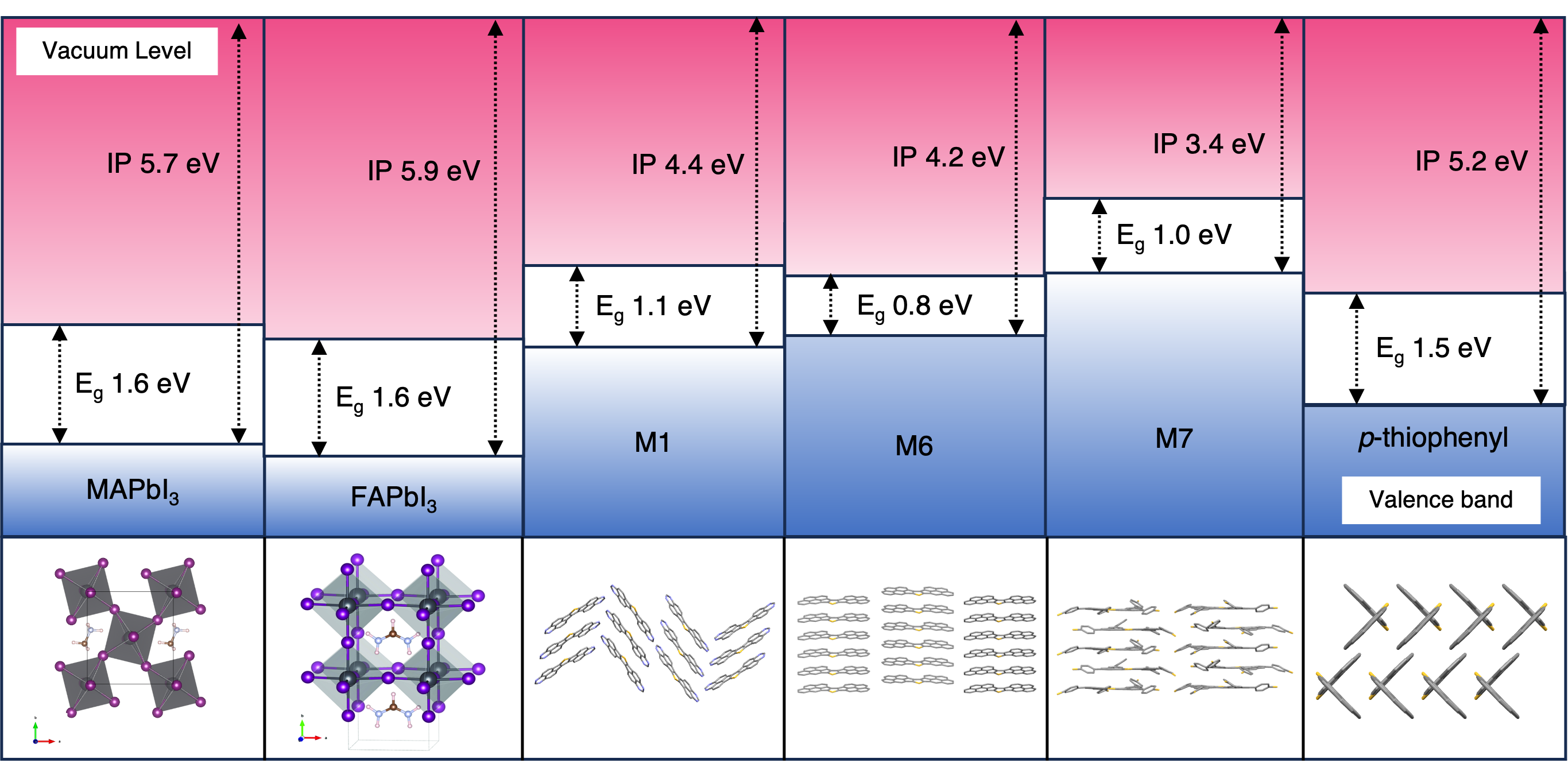}
	\caption{PBE-D3 calculated vacuum-aligned band edges for the top functionalized candidates compared to MAPbI$_3$ and FAPbI$_3$ perovskites. Computed band gaps (E$_g$ in eV) and ionization potentials (IPs in eV) are also given.}
	\label{electronic_top}
\end{figure*}